\documentclass[prl,reprint,twocolumn,showpacs,superscriptaddress,floatfix,aps,10pt]{revtex4-2}

\usepackage{amsmath,amsthm,amssymb}
\usepackage{dcolumn,bm,hyperref}
\usepackage{graphicx,subfigure,verbatim}
\usepackage{txfonts}    
\usepackage{newtxtext}
\usepackage{xcolor}
\usepackage{soul}

\renewcommand{\vec}[1]{{\bm{\mathrm{#1}}}}
\newcommand{\vhat}[1]{\hat{\bm{\mathrm{#1}}}}
\let\Re\undefined
\let\Im\undefined
\DeclareMathOperator{\Im}{Im}
\DeclareMathOperator{\Re}{Re}

\usepackage{xcolor}

\newcommand{\delete}[1]{}

\begin{document}

\title{Orbital Pumping Incorporating Both Orbital Angular Momentum and Position
}
\author{Seungyun Han$^{**}$}
\affiliation{Department of Physics, Pohang University of Science and Technology, Pohang 37673, Korea}
\author{Hye-Won Ko$^{**}$}
\affiliation{Department of Physics, Korea Advanced Institute of Science and Technology, Daejeon 34141, Korea}
\author{Jung Hyun Oh}
\affiliation{Department of Physics, Korea Advanced Institute of Science and Technology, Daejeon 34141, Korea}
\author{Hyun-Woo Lee}
\email{hwl@postech.ac.kr}
\affiliation{Department of Physics, Pohang University of Science and Technology, Pohang 37673, Korea}
\author{Kyung-Jin Lee}
\email{kjlee@kaist.ac.kr}
\affiliation{Department of Physics, Korea Advanced Institute of Science and Technology, Daejeon 34141, Korea}
\author{Kyoung-Whan Kim}
\email{kwkim@yonsei.ac.kr}
\affiliation{Department of Physics, Yonsei University, Seoul 03722, Korea}
\affiliation{Center for Spintronics, Korea Institute of Science and Technology, Seoul 02792, Korea}

\begin{abstract}
We develop a theory of adiabatic orbital pumping, highlighting qualitative differences from spin pumping. An oscillating magnetic field pumps not only orbital angular momentum current but also orbital angular position current. The latter, which has no spin counterpart, underscores the incompleteness of existing orbital torque theories. Importantly, both types of orbital currents can be detected as transverse electric voltages, which contain considerable second harmonic components unlike in spin pumping. Moreover, orbital currents can be pumped by lattice dynamics that carry phonon angular momentum, implying that orbital currents can, in turn, induce phonon angular momentum. Our work open up new possibilities for generating orbital currents and provides a broader understanding of the interplay between spin, orbital, and phonon dynamics.
\end{abstract}

\maketitle

{\it Introduction.--} Orbital dynamics~\cite{Go2021,Jo2024} in solids has recently attracted considerable interest due to its potential to enhance the functionalities of spintronic devices by exploiting the orbital angular momentum of electrons. Phenomena such as the orbital Hall effect~\cite{bernevig2005b,tanaka2008,kontani2008,kontani2009,tokatly2010,go2018,jo2018,bhowal2021,cysne2021,Choi2023}, orbital Edelstein effect~\cite{go2017,johansson2021,hamadi2023}, orbital magnetoresistance~\cite{ko2020,ding2022}, and orbital torque~\cite{go2020,zheng2020,tazaki2020,lee2021,kim2021,Sala2022} could, respectively, complement and strengthen the functionalities achieved by the spin Hall effect~\cite{sinova2004,sinova2015}, spin Edelstein effect~\cite{edelstein1990,sanchez2013}, spin magnetoresistance~\cite{nakayama2016}, and spin torque~\cite{Slonczewski1996,Berger1996}. Moreover, stronger orbital-related phenomena than spin-related ones in a wider class of materials~\cite{jo2018,johansson2021,hamadi2023} open the field of orbitronics.

Despite recent progress, the fundamental understanding of orbital dynamics lags far behind that of spin dynamics. Advancing orbitronics requires identifying qualitative differences between orbital and spin dynamics. In this regard, we note that orbitals have more degrees of freedom than spin. Whereas three spin operators ($S_x,S_y,S_z$) completely describe spin 1/2 states, three orbital angular momentum (OAM) operators ($L_x,L_y,L_z$) are insufficient for describing general orbital states because the size of density matrix for orbital is larger than that of spin. For instance, the orbital direction $\phi$ of the real $p$-orbital state $\cos\phi|p_x\rangle+\sin\phi|p_y\rangle$ cannot be specified from its expectation values of the OAM operators, which are zero. This necessitates additional orbital operators, called orbital angular position (OAP) operators~\cite{han2022}, defined by symmetric combinations of the OAM operators for $p$ orbitals~\cite{han2022}. Whereas the OAM has many properties in common with spin, the OAP does not have any spin counterpart. Thus, the OAP may hold the key for the further development of orbitronics beyond spintronics. A specific Hall response of the OAP was theoretically predicted~\cite{han2022}, i.e., the orbital-torsion Hall effect that refers to the OAP-dependent transverse electron flow caused by an electric field.
However, the broader responses of  OAP and their implications to orbitronics remain poorly understood.

Adiabatic pumping~\cite{Thouless1983} provides critical insights into fundamental dynamics~\cite{Citro2023}. Adiabatic spin pumping has significantly advanced our understanding of magnetic damping enhancement~\cite{tserkovnyak2002}, spin motive force~\cite{kim2012,tatara2013,saslow2007,cheng2014,chen2015}, and spin torque~\cite{Tserkovnyak2005}. Similarly, adiabatic orbital pumping (or orbital pumping in short) offers a powerful means to explore the fundamentals of orbital dynamics. However, recent studies on orbital pumping~\cite{santos2023,hayashi2023, go2023orbital} have been limited to OAM and have primarily focused on similarities to spin pumping, leaving key differences unexplored.
%
\begin{figure}[t]
\includegraphics[scale=0.125]{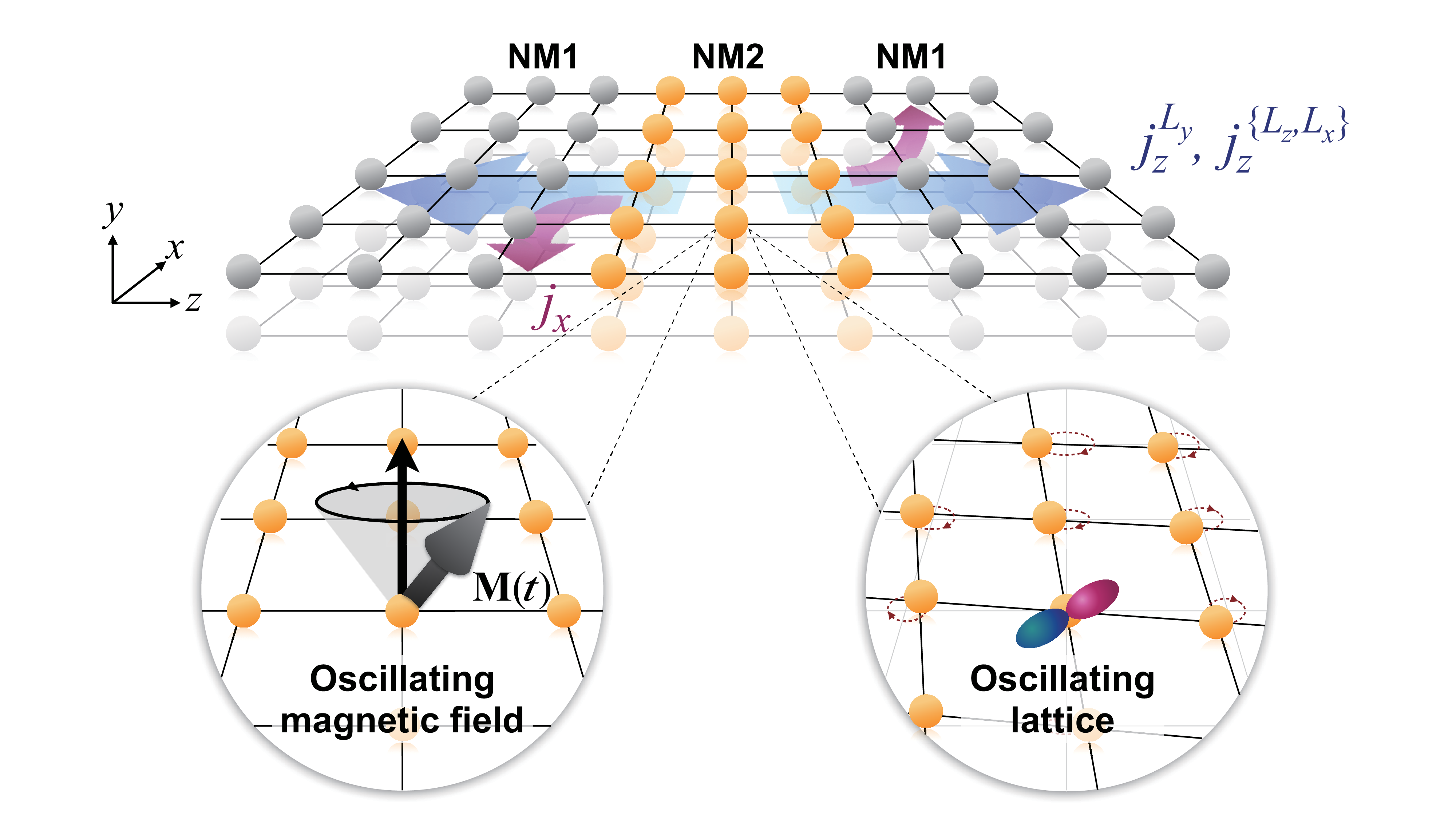}
\caption{
Schematic illustration of orbital pumping in a model system, NM1/NM2/NM1  (NM = normal metal), driven by an oscillating magnetic field or vibrating lattice. For the magnetic-field-driven case, NM2 may be considered a ferromagnet. $j_z^i$ is an orbital current flowing along $z$ with orbital $i$, and $j_x$ is a charge current flowing along $x$.
}
\label{fig:1}
\end{figure}

In this Letter, we present a theory of orbital pumping that incorporates both OAM and OAP degrees of freedom. Our first main finding is that an oscillating magnetic field applied to a ferromagnet (FM) or normal metal (NM) pumps both OAM and OAP currents into neighboring materials (Fig.~\ref{fig:1}). Onsager reciprocity suggests that orbital torque can be generated in FM not only by OAM injection but also by OAP injection (via orbital-torsion Hall effect~\cite{han2022}). The latter mechanism for generating orbital torque has not been considered in current theories~\cite{go2020,go2020b} and may demand a re-examination of previous orbital torque experiments~\cite{zheng2020,tazaki2020,lee2021,kim2021,Sala2022}.
Our second finding is that OAM and OAP currents can also be pumped by lattice dynamics carrying phonon angular momentum (PAM). Onsager reciprocity implies that PAM can be excited by injecting electron orbital currents, even in non-chiral materials. Considering recent interest in PAM~\cite{zhang2014,zhu2018}, this finding will stimulate further studies on the connection between electron's orbital and PAM. It in turn necessitates studies on electron-phonon interaction to include orbital information transfer to phonons.
Additionally, we show that orbital pumping generates a second-harmonic response, even in minimally nonlinear situations, which contrasts with the DC and first-harmonic signals produced by spin pumping. This offers a clear experimental method to \textit{qualitatively} distinguish the orbital dynamics from the spin dynamics, providing a means to address the ongoing debate whether existing measurements of orbital relaxation might have actually measured spin relaxation~\cite{Rang2024}.  

{\it Orbital pumping by oscillating magnetic field.--}
In a FM, the perturbation ${\cal H}(t)=J_{\rm ex}\vec{L}\cdot \vec{M}(t)$ may induce orbital pumping, where $\vec{L}$ is the (dimensionless) OAM operator, $J_{\rm ex}$ is the coupling strength, and ${\bf M}(t)$ is the unit vector along the magnetization, whose direction is controlled by a magnetic field. Alternatively, orbital pumping may occur from a NM (Fig.~\ref{fig:1}) when the field couples directly to $\vec{L}$. In this case, $\vec{M}(t)$ in ${\cal H}(t)$ may be interpreted as the unit vector along the field direciton. We ignore the spin degree of freedom to focus solely on orbital responses. To get a tractable analytic formula, we consider a $p$-orbital system as a minimal model 
and adopt the atom-center approximation, treating the OAM operator $\vec{L}$ as $3\times3$ matrices in $p$-orbital space. We also assume an ideal case where three $p$-orbitals are degenerate in equilibrium. This ideal case reveals critical qualitative differences between orbital pumping and spin pumping. As shown later, our numerical calculations confirm that the predictions from this ideal case persist in real situations, where the $p$-orbital degeneracy is lifted. We construct a model system of NM1/NM2/NM1 structure (Fig.~\ref{fig:1}) and derive orbital pumping currents induced by time-dependent perturbations to NM2, where $J_{\rm ex}$ in ${\cal H}(t)$ is finite. The $3\times 3$ matrix Green's function $g$ provides a complete description of single-particle dynamics. We abbreviate the explicit position dependence of $g$ for simplicity. For a degenerate $p$-orbital system, the $\vec{L}\cdot\vec{M}(t)$ is conserved and $g$ is expanded as 
\(
    g=\sum_{m=-1,0,1} g_m P_m(t),
\) 
where $g_m$ and $P_m(t)=|\vec{L}\cdot\vec{M}(t)=m\rangle \langle \vec{L}\cdot\vec{M}(t)=m|$ are, respectively, the Green's function and projection operator associated with an OAM eigenstate having an eigenvalue $m$ of $\vec{L}\cdot\vec{M}(t)$.

The pumped current $j_\alpha$, flowing along the direction $\alpha$, is given by a $3\times 3$ matrix, similarly to the $2\times 2$ current for spin pumping~\cite{tserkovnyak2002}. Thus, $j_\alpha$ may be expanded in terms of the OAM and OAP as follows,
\begin{equation}
j_\alpha/(-e)=\vec{j}_\alpha^{\rm OAM}\cdot\vec{L}+\sum_{\beta\gamma}j_{\alpha,\beta\gamma}^{\rm OAP}\{L_\beta,L_\gamma\},
\label{general_pumped_current}
\end{equation}
since the three $3\times 3$ OAM operators $L_x$, $L_y$, $L_z$ and the six $3\times 3$ OAP operators $\{L_\beta,L_\gamma\}$ form a complete set of basis for general $3\times 3$ hermitian matrices~\cite{han2022}. Here $\vec{j}_\alpha^{\rm OAM}$ and $j_{\alpha,\beta\gamma}^{\rm OAP}$ amount to the pumped OAM and OAP currents, respectively, flowing along $\alpha$ direction. Thus, the complete description of the orbital degrees of freedom naturally leads to the prediction that not only OAM but also OAP are pumped, which is our first main finding. 
Employing the method developed in Ref.~\cite{chen2015}, we obtain
\begin{equation}
    \textbf{j}^{\text{OAM}}_\alpha = \frac{1}{4\pi}\Re\left[\left(\frac{G^{1,0}_\alpha+G^{0,-1}_\alpha}{2}\right)\left(\textbf{M}\times\frac{d\textbf{M}}{dt}-i\frac{d\textbf{M}}{dt}\right)\right],\label{OAM_1}
\end{equation}
\begin{eqnarray}
    j^{\text{OAP}}_{\alpha, \beta\gamma} &=&\frac{1}{8\pi}\Re\left[\left(\frac{G^{1,0}_\alpha-G^{0,-1}_\alpha}{2}\right)\right.\nonumber\\
    &&\quad\times\left.\left(M_\beta\left(\textbf{M}\times \frac{d\textbf{M}}{dt}\right)_\gamma-iM_\beta\frac{dM_\gamma}{dt}+(\beta\leftrightarrow \gamma) \right)\right], \label{OAP_1}
\end{eqnarray}
where
\begin{equation}
    G^{\mu,\nu}_\alpha = \frac{J_{\text{ex}}\hbar^2}{m_e}\int dr' [g^R_\mu (r,r') \overset{\leftrightarrow}{\partial_\alpha} g^A_\nu (r',r)] \label{OMC},
\end{equation}
which is the orbital-generalization of the spin-mixing conductance for the spin pumping~\cite{tserkovnyak2002}. Here, $g^{R/A}$ represents retarded/advanced Green's function of the NM1/NM2/NM1 heterostructure, $\overset{\leftrightarrow}{\partial_\alpha}$ is the antisymmetric differential operator, and $\int dr'$ denotes the volume integral. 

To investigate implications of Eqs.~(\ref{OAM_1}) and (\ref{OAP_1}), we consider a specific situation in Fig.~\ref{fig:1}, where $\alpha=z$ is the pumping direction. When ${\bf M}(t)$ rotates in the $zx$ plane (${\bf M}(t)=\vhat{z}\cos\omega t+\vhat{x}\sin\omega t$), $j_z^{\rm OAM} = \vec{j}_z^{\rm OAM}\cdot\vec{L}$ and $j_z^{\rm OAP} = \sum_{\beta\gamma}j_{z,\beta\gamma}^{\rm OAP}\{L_\beta,L_\gamma\}$ become
\begin{equation}
    {j}^{\text{OAM}}_z = \frac{\omega}{4\pi}\left\{ \Re[G_L^+]L_y+\Im[G_L^+](L_x \cos\omega t-L_z \sin\omega t)\right\},\label{OAM_2}
\end{equation}
\begin{eqnarray}
    {j}^{\text{OAP}}_{z} &=&\frac{\omega}{4\pi}\Im[G_L^-]\left(\left\{L_z,L_x\right\}\cos2\omega t-(L_z^2-L_x^2)\sin2\omega t\right)\nonumber\\
    &+&\frac{\omega}{4\pi}\Re[G_L^-]\left(\left\{L_y,L_z\right\}\cos\omega t+\left\{L_x,L_y\right\}\sin\omega t\right), \label{OAP_2}
\end{eqnarray}
where $G_L^\pm=(G^{1,0}_z \pm G^{0,-1}_z)/2$. Here, 
$j_z^{\rm OAM}$ and $j_z^{\rm OAP}$ are the operator expressions of OAM and OAP currents, respectively. 
According to Eq.~(\ref{OAM_2}), the OAM pumping consists of a DC component carrying $L_y$ and first-harmonic ($\omega$) ones carrying $L_x,L_z$, which is analogous to the spin pumping. On the other hand, the OAP pumping [Eq.~(\ref{OAP_2})] consists of the first-harmonic components carrying $\{L_y,L_z\},\{L_x,L_y\}$ and the {\it second}-harmonic ($2\omega$) ones carrying $\{L_z,L_x\}$ and $L_z^2-L_x^2$. This difference may be utilized to distinguish the orbital pumping from the spin pumping. As a side remark, the second-harmonic signal generated by orbital pumping is a linear response, which is conceptually different from nonlinear phenomena like the Suhl instability~\cite{anderson1955,suhl1957,suhl1958}, and thus experimentally distinguishable.
\delete{, which we use to show the dynamics of each orbital degree of freedom explicitly. Equations~(\ref{OAM_2}) and (\ref{OAP_2}) predict that orbital pumping currents in ideal situations (i.e., no crystal field spitting) consist of a DC component carrying $L_y$, first-harmonic ($\omega$) ones carrying $L_x,L_z,\{L_y,L_z\}$, and $\{L_x,L_y\}$, and second-harmonic ($2\omega$) ones carrying $\{L_z,L_x\}$ and $(L_z^2-L_x^2)$.}

Next, we numerically examine orbital pumping in a more realistic situation in the sense that the degeneracy of three $p$-orbitals is lifted by crystal fields. We adopt a $sp3$ tight-binding model in a simple cubic lattice with nearest-neighbor hopping. 
We assume that all energy bands near the Fermi energy have $p$-character, so the model describes a $p$-orbital system. Still, the introduction of an $s$-character band located above the Fermi energy allows for the generalization of the model to a more realistic situation, where real orbital eigenstates vary with crystal momentum (orbital texture) through $sp$ hybridization~\cite{ko2020,Han2023CAP}.
See \cite{supple} for further details of the model.
\begin{figure}[t]
\includegraphics[scale=0.45]{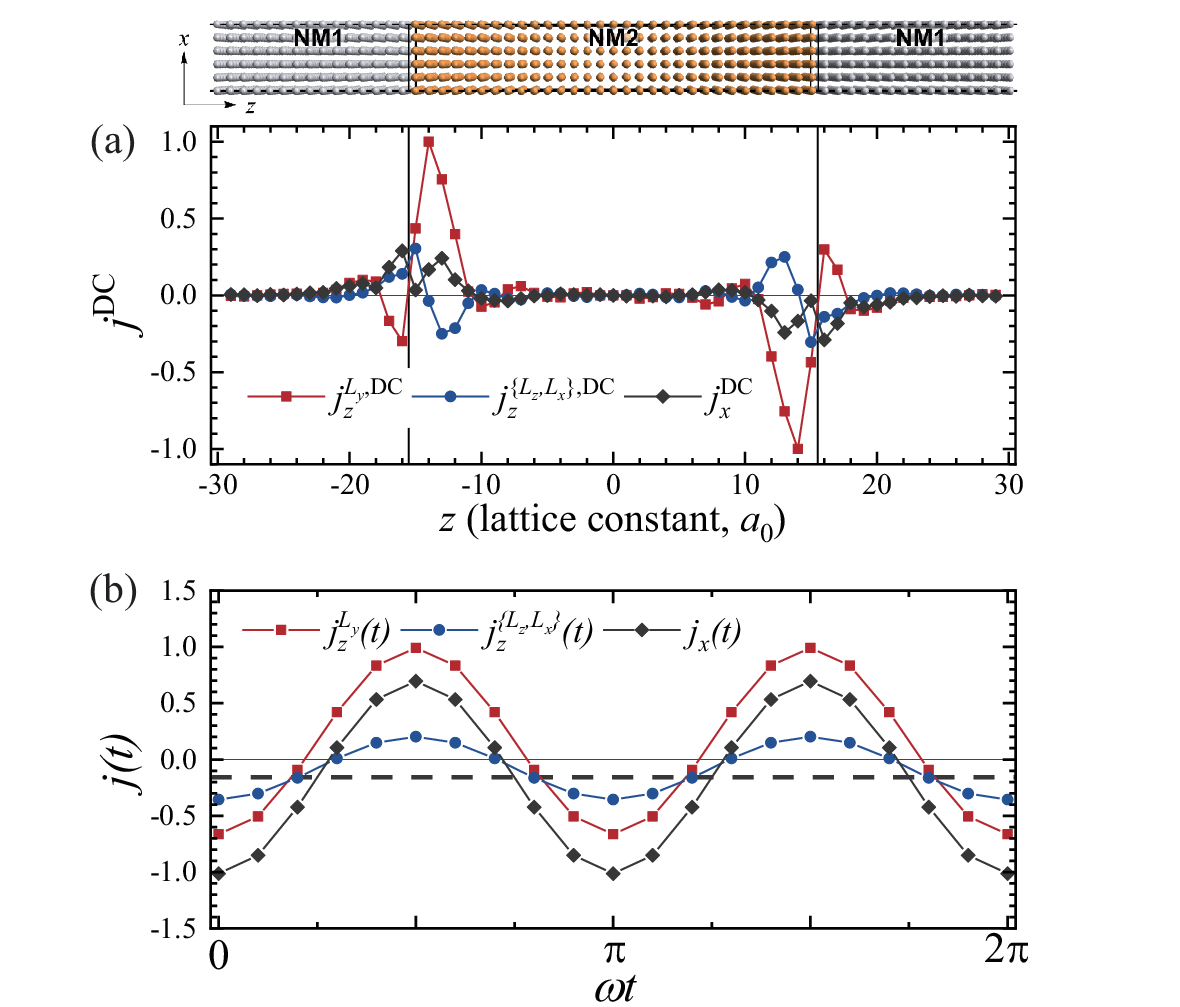}
\caption{ 
(a) Spatial profile of DC pumping currents, $j_z^{L_y,\rm DC}$ (red), $j_z^{\{L_z,L_x\},\rm DC}$ (blue), and $j_x^{\rm DC}$ (gray), driven by the time-dependent magnetic field, ${\bf L}\cdot{\bf M}(t)$ for ${\bf M}(t)=\vhat{z}\cos\omega t+\vhat{x}\sin\omega t$, on NM2 layer. (b) Temporal dependence of pumping currents $j(t)$ at the right interface ($z=16~a_0$). The thick dashed horizontal line in (b) shows the DC component of transverse charge current $j_x(t)$.
}
\label{fig:2}
\end{figure}

For the same $\vec{M}(t)$ given above, we calculate numerically the pumped orbital current using the linear response theory in the adiabatic limit (see \cite{supple}). The calculation confirms that all of the orbital currents predicted by Eqs.~(\ref{OAM_2}) and (\ref{OAP_2}) are 
pumped by the ${\bf M}(t)$ oscillation. Figure~\ref{fig:2} presents some of the numerical results: DC OAM current ($j_z^{L_y,{\rm DC}}$) [Fig.~\ref{fig:2}(a)] 
and second-harmonic OAP current ($j_z^{\{L_z,L_x\}}(t)$) [Fig.~\ref{fig:2}(b)]. It is worth mentioning that OAP pumping currents are comparable in magnitude to OAM pumping currents (Fig.~\ref{fig:2}). Onsager reciprocity then implies that the orbital torque by OAP injection may be comparable to the conventional orbital torque by OAM injection. In addition, we find pumping of orbital currents that are absent in Eqs.~(\ref{OAM_2}) and (\ref{OAP_2}). For instance, the OAM (OAP) current has an unexpected second-harmonic (DC) component [Fig.~\ref{fig:2}(b)]. This is a result of the inclusion of the crystal field splitting, which can convert OAM current to OAP current and vice versa~\cite{han2022}. 

In addition to the interconversion between OAM and OAP, we find \textit{intra}conversion among OAM components and \textit{intra}conversion among OAP components, which can be called OAM and OAP swapping effects, respectively (see \cite{supple} for details). The OAM swapping effect is analogous to the spin swapping effect~\cite{lifshits2009,sadjina2012,saidaoui2016}, 
in line with a recent theory~\cite{Manchon2023}. The OAP swapping effect, however, does not have any spin counterpart. The OAM and OAP swapping effects require the orbital texture but do not require the spin-orbit coupling.


The short decay length ($\leq$ 1 nm) of orbital currents in NM1 [Fig. \ref{fig:2}(a)] is consistent with recent theoretical calculations~\cite{urazhdin2023,belashchenko2023,Rang2024}, but much shorter than experimental values (5 $\sim$ 100 nm)~\cite{lee2021,Choi2023,moriya2024,Otani2024}. Possible reasons for this discrepancy include disorder scattering~\cite{Liu2024,Sohn2024}, multi-grain effects~\cite{liao2022,idrobo2024}, and electron-phonon scattering~\cite{Otani2024}, all of which go beyond the scope of this paper. We argue that this controversy does not affect the predicted qualitative features. In particular, even if the orbital relaxation length is very short, the pumped OAM and OAP currents can be electrically measured through their conversion to transverse charge currents [Fig. \ref{fig:2}(a)] via the inverse orbital Hall effect~\cite{han2022,Wang2023,Xu2023} and the inverse orbital-torsion Hall effect~\cite{han2022}, respectively, just as pumped spin currents are electrically measured even when their decay length is very short ($\sim$ 1 nm) in materials with strong SOC.

A recent orbital pumping experiment~\cite{hayashi2023} reported a DC charge current and attributed it entirely to conversion from pumped DC OAM current~\cite{go2023orbital}. However, according to our calculation, the measured DC charge current contains an additional contribution due to conversion from a pumped DC OAP current. The converted charge current also contains a second-harmonic component since pumped OAM and OAP currents contain second-harmonic components [Fig.~\ref{fig:2}(b)]. The second-harmonic charge current remains unexplored experimentally. It is worth noting that the generalized pumping equation derived in Supplemental Materials (SM)~\cite{supple} indicates that a $d$-orbital system may also produce a fourth-harmonic component. As a spin pumping current contains only DC and first-harmonic components~\cite{tserkovnyak2002}, higher-harmonics pumping signals are
a unique feature of orbital pumping. 
%
%
%
\begin{figure}[t]
\begin{center}
\includegraphics[scale=0.45]{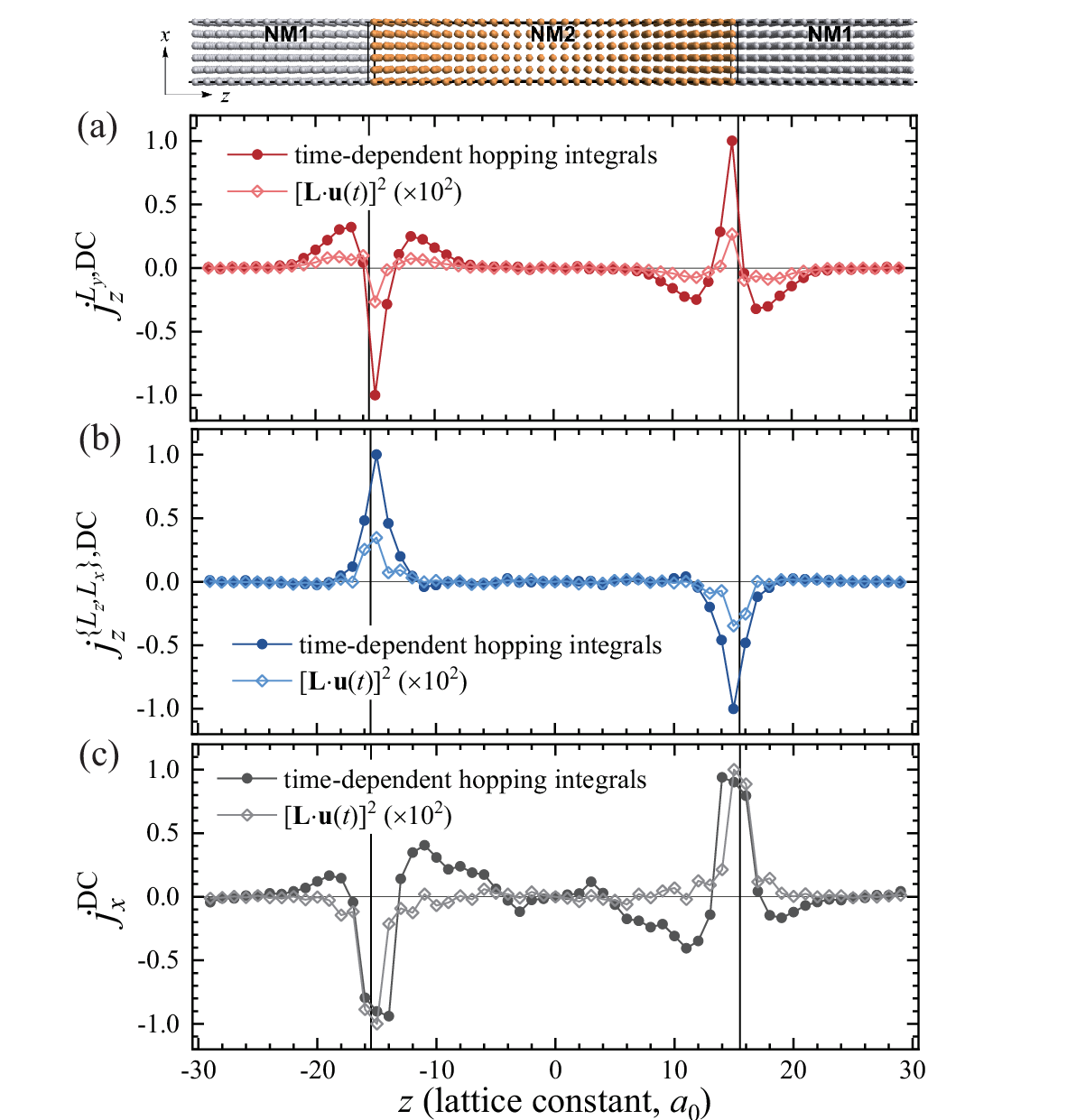}
\caption{
Spatial profile of DC pumping currents (a) $j_z^{L_y,\rm DC}$, (b) $j_z^{\{L_z,L_x\},\rm DC}$, and (c) $j_x^{\rm DC}$ driven by the lattice dynamics, which is imposed by time-dependent variations of tight-binding hopping integrals (solid circles) and $[{\bf L}\cdot{\bf u}(t)]^2$ for ${\bf u}(t)=\vhat{z}\cos\omega t+\vhat{x}\sin\omega t$ (open diamonds).
}
\label{fig:3}
\end{center}
\end{figure}

{\it Orbital pumping by lattice dynamics.--} Next, we explore orbital pumping due to lattice dynamics, which can be realized by a slowly time-varying strain. The resulting adiabatic variation of a crystal may cause: variations of i) bonding lengths and ii) bonding angles. Each variation manifests in the tight-binding model as corrections to hopping integrals and directional cosines, respectively. The hopping integrals are assumed to follow a power law of bonding length~\cite{froyen1979,harrison1980}. We integrate the lattice dynamics driven by adiabatically varying strains into our model and numerically calculate the resulting orbital pumping (see SM~\cite{supple} for details). 

We consider a biaxial strain applied to the NM2 layer in our NM1/NM2/NM1 trilayer system (Fig.~\ref{fig:1}). The biaxial principal axes of the strain are assumed to rotate slowly in the $zx$ plane, which makes our $sp$-implementedcubic lattice undergo a circularly rotating deformation. 
This situation mimics the excitation of a long-wavelength phonon vibration with PAM. Numerical calculations with the time-dependent hopping integrals show that both DC OAM current $j_z^{L_y,{\rm DC}}$ [Fig.~\ref{fig:3}(a)] and DC OAP current $j_z^{\{L_z,L_x\}{\rm DC}}$ [Fig.~\ref{fig:3}(b)] are pumped to the neighboring NM1 layers, and converted to DC transverse charge current $j_z^{\rm DC}$ [Fig.~\ref{fig:3}(c)]. Thus, the lattice dynamics with PAM generates similar orbital pumping as an oscillating magnetic field does. This is our second main finding, which illustrates a fundamental channel of angular momentum transfer between electron orbital degrees of freedom and phonons, in addition to the well-known channel between electron spin degree of freedom and magnetization. The OAM pumping induced by the lattice dynamics is the reverse process of crystal field torque~\cite{go2020b}, where electron OAM, generated by external perturbations such as an electric field, is absorbed by the lattice. Our calculation implies that the reverse process of the OAP current offer an additional mechanism of the crystal field torque that has not been identified before.

To gain an insight, we consider a simplified model focusing on the orbital splitting. Since this perturbation is time-reversal even, the corresponding perturbation Hamiltonian should be expressed in terms of even-order products of the OAM operators, which are OAP operators.
To be more specific, we consider a strain applied along the $\vec{u}$ direction, and model the resulting perturbation as $K[{\bf L}\cdot{\bf u}(t)]^2$ where $K$ is the strain-OAP coupling strength (see SM~\cite{supple}). This modeling is based on the reasoning that the strain tends to make the orbital energy of the $p_u$ orbital different from those of the other $p$ orbitals perpendicular to it. The perturbation $K[{\bf L}\cdot{\bf u}(t)]^2$ does exactly this job. The open diamonds in Fig.~\ref{fig:3} show the orbital currents pumped by the $K[{\bf L}\cdot{\bf u}(t)]^2$ perturbation incorporated to the numerical model. Note that the results from this simplified model \delete{can} reproduce the characteristics of orbital pumping obtained from the previous numerical calculation that incorporates i) and ii). 

We simplify the model further so that analytic calculation becomes possible. In the limit of the negligible crystal field, the orbital splitting is governed by the perturbation $K[{\bf L}\cdot{\bf u}(t)]^2$ and the Green's function in the presence of the perturbation can be expressed as $g= g_0 I + (g_1 -g_0) [\textbf{L}\cdot \textbf{u}(t)]^2$. Then, we obtain
\begin{align}
	\textbf{j}^{\text{OAM}}_\alpha&= \frac{1}{4\pi}\Re[G^{1,0}_\alpha] \textbf{u}\times \frac{d\textbf{u}}{dt},\label{OAM_3} \\
	j^{\text{OAP}}_{\alpha,\beta\gamma}&= \frac{1}{8\pi}\Im[G^{1,0}_\alpha] \frac{d(u_\beta u_\gamma)}{dt},\label{OAP_3} 
\end{align}
where $G_\alpha^{\mu,\nu}$ is given by Eq.~(\ref{OMC}) except that $J_{\rm ex}$ is replaced by $K$. We take $\vec{u}(t)=\hat{\vec{z}}\cos \omega t + \hat{\vec{x}}\sin \omega t$ to mimic the adiabatically rotationg biaxial strain examined above numerically. Equation~(\ref{OAM_3}) then predicts the pumping of the DC OAM current $j^{L_y,{\rm DC}}_z$, which can be understood as the transfer of PAM to electron OAM~\cite{yao2022}. On the other hand, the total derivative in Eq.~(\ref{OAP_3}) predicts the pumped OAP current $j^{\{L_z,L_x\}}_{z}(t)$ not to have any DC component. The emergence of DC OAP current $j^{\{L_z,L_x\},{\rm DC}}_{z}$ in numerical calculations [Fig.~\ref{fig:3}(b)]) is again attributed to the orbital texture, resulting in an interconversion between OAM and OAP, as generally proven in Ref.~\cite{han2022}. 

{\it Discussion and outlook.--} We demonstrated orbital pumping driven by either  
oscillating ${\bf M}(t)$ or lattice dynamics ${\bf u}(t)$. The simultaneous emergence of OAM and OAP pumping for both pumping methods highlights the necessity of considering both orbital degrees of freedom when describing orbital dynamics. This result has an important implication for orbital torque. It has been believed that orbital torque originates solely from the injection of OAM current~\cite{go2020}. However, the pumping of OAP current through magnetization dynamics suggests the existence of its inverse process: the generation of magnetic torque through the injection of OAP current. This previously unrecognized inverse process, which we term ``OAP torque", introduces a new dimension to the understanding of orbital torque. For instance, the OAP torque implies that the Onsager reciprocity between orbital pumping and orbital torque is validated only when one considers both OAM and OAP contributions. 

Our theory of orbital pumping also offers an exploitable method to generate orbital currents, just as spin pumping serves as an efficient means to generate a pure spin current~\cite{mosendz2010,ando2011}. Considering that our theory neglects the spin degree of freedom, we expect that orbital magnets may be ideal materials to generate an almost pure orbital current through pumping since their magnetization is dominated by the orbital magnetic moment. Examples include CoMnO$_3$ films~\cite{koizumi2019}. We note that previous suggestions~\cite{han2022} for the orbital current generation method rely on twisted heterostructures, which can be challenging to implement, or on low-gap semiconductors with limited availability of the required materials. In contrast, the high-harmonics pumping driven by the OAP degree of freedom can be realized with various materials since spin pumping is governed by the DC and first-harmonic components. Thus, orbital pumping extends the scope of orbitronics to a broader range of systems. We also note that this property stems from the distinct behaviors of high-order products of $L$ operators with respect to rotational transformations. Thus, it is a general property not limited to
$p$-orbital systems.

Furthermore, we argue that orbital pumping is not limited to multilayer structures and may be extended to {\it single-layer bulk materials}. Recalling that the spin motive force is a continuum version of spin pumping~\cite{zhang2009} and occurs in single-layer bulk materials, we anticipate the presence of an orbital motive force (continuum version of orbital pumping) when a single-layer system exhibits an inhomogeneous crystal field. Thus, the orbital motive force would encompass OAP contributions, which lack their spin counterparts. Notably, the physical properties of the orbital motive force would differ from those of the spin motive force due to the presence of orbital texture, a factor not accounted for in the theory of the spin motive force. The exploration of this topic remains a subject for future work.
 
Lastly, our work sheds light on the transfer of angular momentum between electrons and phonons. Theoretical estimations~\cite{zhang2014,zhang2015,ren2021} and experimental measurements~\cite{zhu2018,holanda2018,sasaki2021} illustrate that PAM can have a considerable magnitude contrary to early assumptions and plays a nontrivial role in various phenomena such as magnetization relaxation~\cite{streib2018} and ultrafast demagnetization~\cite{dornes2019,tauchert2022}. Intriguingly, recent many-body treatment~\cite{mentink2019} shows that a complete picture of angular momentum transfer between electron and phonon subsystems requires OAM as a key ingredient since the electron-phonon coupling is independent of spin. The strain-induced orbital pumping weighs heavily on this connection of orbital and lattice and call for a wider viewpoint on PAM dynamics~\cite{hamada2020,yao2022,yao2023,juraschek2019,juraschek2020} assisted by the orbital degree of freedom~\cite{xiao2021}.


\begin{acknowledgments}
H.-W.L. acknowledges Antonio Azevedo for the fruitful discussion. This work was supported by the National Research Foundation of Korea (NRF) funded by the Ministry of Science and ICT (2020R1A2C3013302, 2022M3I7A2079267, RS-2024-00334933, RS-2024-00410027) and the KIST Institutional Program. S.H. and H.-W.L. were supported by the Samsung Science and Technology Foundation (BA-1501-51).
\end{acknowledgments}

$^{**}$S.H. and H.-W.K. contributed equally to this work.

\end{document}


\title{Supplemental Material for "Orbital Pumping Incorporating Both Orbital Angular Momentum and Position"
}
\author{Seungyun Han$^{**}$}
\affiliation{Department of Physics, Pohang University of Science and Technology, Pohang 37673, Korea}
\author{Hye-Won Ko$^{**}$}
\affiliation{Department of Physics, Korea Advanced Institute of Science and Technology, Daejeon 34141, Korea}
\author{Jung Hyun Oh}
\affiliation{Department of Physics, Korea Advanced Institute of Science and Technology, Daejeon 34141, Korea}
\author{Hyun-Woo Lee}
\email{hwl@postech.ac.kr}
\affiliation{Department of Physics, Pohang University of Science and Technology, Pohang 37673, Korea}
\author{Kyung-Jin Lee}
\email{kjlee@kaist.ac.kr}
\affiliation{Department of Physics, Korea Advanced Institute of Science and Technology, Daejeon 34141, Korea}
\author{Kyoung-Whan Kim}
\email{kwkim@yonsei.ac.kr}
\affiliation{Department of Physics, Yonsei University, Seoul 03722, Korea}
\affiliation{Center for Spintronics, Korea Institute of Science and Technology, Seoul 02792, Korea}

\maketitle

\section{Introduction to orbital angular position}

\begin{figure}[b]
\begin{center}
\includegraphics[scale=0.18]{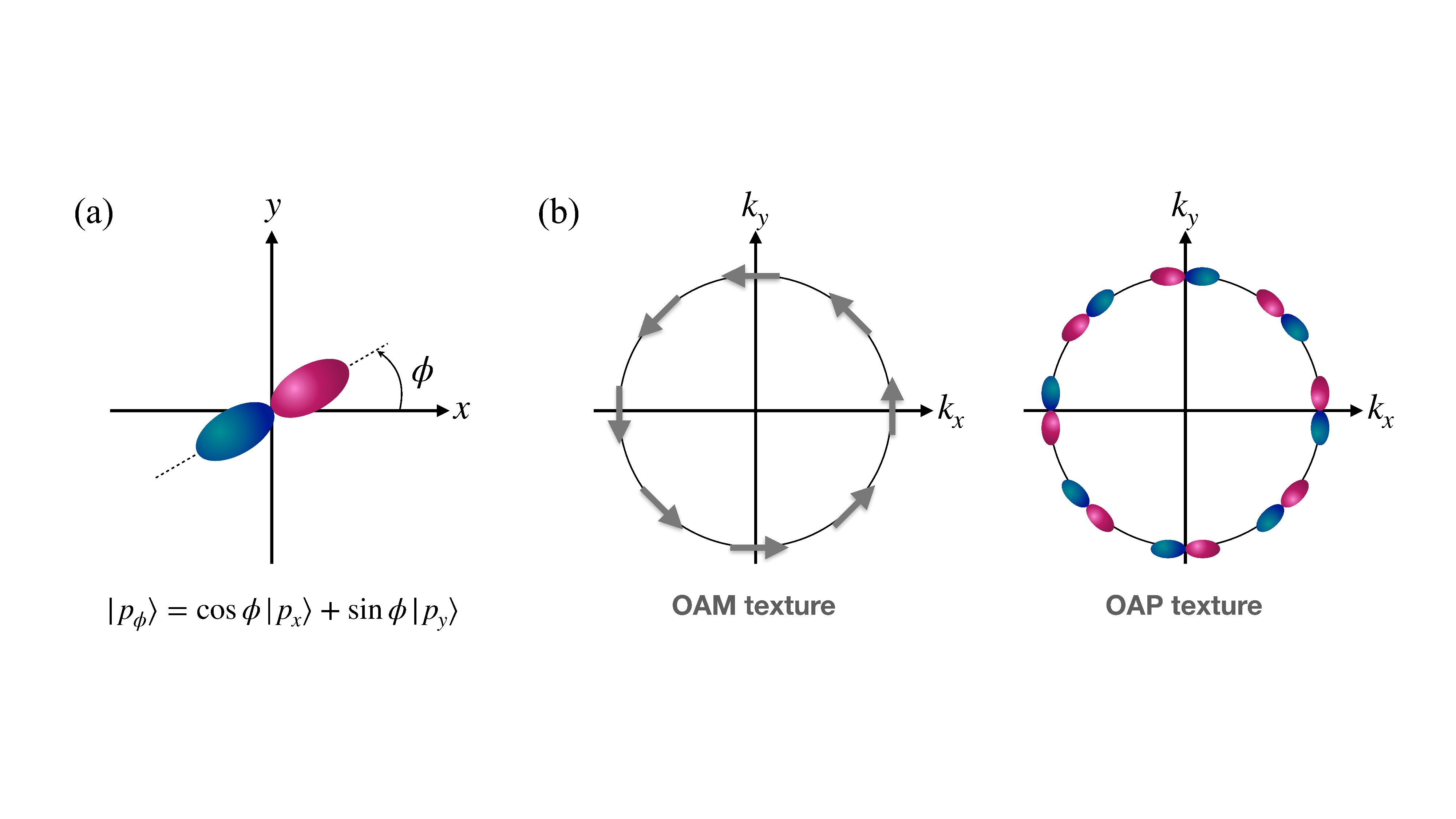}
\caption{(a) Schematic of an arbitrary $p$-orbital state $|p_\phi\rangle$ in $xy$ plane. (b) Illustration of (left) OAM texture, such as the orbital Rashba effect, and (right) tangential OAP texture in momentum space. Gray arrow indicates the OAM vector at each crystal momentum.
}
\label{fig:s1}
\end{center}
\end{figure}

In this section, we outline the comprehensive description of orbital properties including both orbital angular momentum (OAM) and orbital angular position (OAP), drawn from Ref.~\cite{han2022}.

To frame our discussion, let's first consider the well-understood example of spin. A spin system, characterized by an angular momentum quantum number $l=1/2$, is fully encapsulated by the well-known Pauli matrices. These three $2\times2$ matrices effectively describe the orientation of the spin.
In an analogous manner, the orbital degrees of freedom were typically described by the OAM operators, $L_x$,$L_y$, and $L_z$, which stimulated previous studies in terms of OAM, e.g., the orbital Hall effect. 
Contrastingly, orbitals corresponding to $l\ge1$ boast a much richer structure, encapsulating $(2l+1)^2$ degrees of freedom. This complexity extends beyond what can be represented by just three angular momentum operators, imbuing orbitals with additional degrees of freedom and nuances in their directional properties.

For a more tangible illustration, consider a $p$-orbital system where $l=1$. To fully characterize this system, nine degrees of freedom are necessary. This includes six $\{L_i,L_j\}$ operators $(i,j=x,y,z)$ alongside three OAM operators. The $\{L_i,L_j\}$ operators, representing even-order products of the angular momentum operator $L$, possess distinct characteristics such as time-reversal symmetry.
The OAP operators are further classified into two categories. One type of OAP operators $(i=j)$ measures a projection of wave function on $p_i$ state since $\{L_i,L_i\}/2\hbar^2=1-|p_i\rangle\langle p_i|$. The other type of OAP operators $(i\ne j)$ quantifies how much a wave function is rotated with respect to $\hat{i}$ axis or $\hat{j}$ axis in $ij$ plane. For example, consider an orbital state $|p_\phi\rangle=\cos\phi|p_x\rangle+\sin\phi|p_y\rangle$ [Fig.~\ref{fig:s1}(a)] which is constructed based on two orthogonal orbital bases, $|p_x\rangle$ and $|p_y\rangle$. The expectation value of OAP $\{L_x,L_y\}$ is given as $\langle p_\phi|\{L_x,L_y\}|p_\phi\rangle=-\sin2\phi$, which presents the torsion away from the basis orbital states. Note that the OAM of state $|p_\phi\rangle$ is zero because the wave function is given as a real superposition of orbital states. For an arbitrary state composed of (complex) linear combination of orbital bases, both OAM and OAP operators are required for a complete description. 
Unlike the spin case, orbitals allow for an additional degree of freedom in their spatial distribution, justifying the natural emergence of quantities like OAP density and OAP current. These features are absent in spin systems and provide a pathway for orbitals to interact more intimately with other external spatial degrees of freedom, such as the lattice itself.

When we examine orbitals with $l>1$, the complexity increases; a greater number of degrees of freedom is necessary, extending beyond the $\{L_i,L_j\}$ operators. These can be captured by symmetrized operator products like ${\rm Symm}[L_i,L_j,L_k]$, e.g., ${\rm Symm}(L_x^2 L_y)=(L_x^2 L_y+L_x L_y L_x+L_y L_x^2)$. Such higher-order products of $L$ are instrumental in defining the spatial characteristics of orbitals and can be categorized as higher-order OAP.

In centrosymmetric systems without spin-orbit coupling, OAM is quenched owing to the crystal field. Direct coupling of OAM and crystal momenta is achieved when either inversion symmetry or time-reversal symmetry is broken. Note that the former condition is commonly satisfied at interfaces (or surfaces) of materials, which results in OAM texture such as the orbital Rashba effect [left panel of Fig.~\ref{fig:s1}(b)]. In contrast, OAP is not nullified even in equilibrium [right panel of Fig.~\ref{fig:s1}(b)] because terms like $k_\alpha k_\beta\{L_i,L_j\}$ $(\alpha,\beta=x,y,z)$ preserve both symmetries. Instead, OAP directly indicates the coupling between orbital degrees of freedom to the lattice and its texture is responsible for diverse orbital-related phenomena~\cite{han2023}. Hence, OAP is indispensable degrees of freedom along with OAM in both mathematical and physical contexts.

\section{General pumping formula for arbitrary orbital systems}

\subsection{Generalized algebra for high-$l$ orbital spaces}

In the case of spin, one can completely describe the dynamics using the Pauli matrix, enabling us to represent arbitrary traceless operators as vectors. Consequently, the pumping formula can also be expressed in vector form. However, in the case of orbitals with angular momentum quantum number $l$, it is imperative to define a vector space of $(2l+1)^2$ dimensions, leading to the necessity of generalizing the dot product and cross product. This section will provide these definitions and outline their essential algebraic properties. In the following section, we will employ these concepts to demonstrate the straightforward generalization of the pumping formula. 

We define a new symbol "$\dot{=}$" to denote the mapping of an $(2l+1)\times (2l+1)$ matrix $A$ to an $(2l+1)^2$-dimensional vector as follows,
\begin{equation}
    \mathbf{A} \ \dot{=}\ \textit{A},
\end{equation}
where $A= A_iL_i$ (Einstein convention), $\text{Tr}[L_iL_j]=2\delta_{ij}$, and $A_i=\text{Tr}[AL_i]/2$. The orthogonality guarantees the uniqueness and existence of the representation. We can generalize the cross product and dot product that map $\mathbb{R}^{n^2}\times \mathbb{R}^{n^2}$ to $\mathbb{R}^{n^2}$ as,
\begin{align}
\bold{A} \odot \bold{B} &\doteq \frac{1}{2} \{A,B\}, \\
\bold{A} \otimes \bold{B} &\doteq \frac{1}{2i}[A,B].
\end{align}

Below, we present some useful properties of generalized dot and cross products.
\subsubsection{Property 1: Bilinearity}

\begin{align}
(\bold{A}+\bold{B})\odot (\bold{C}+\bold{D}) &= \bold{A}\odot \bold{B} + \bold{B}\odot \bold{C} + \bold{A} \odot \bold{D} +\bold{B} \odot \bold{D},\\
(\bold{A}+\bold{B})\otimes (\bold{C}+\bold{D})&= \bold{A}\otimes \bold{B} +\bold{B}\otimes \bold{C} + \bold{A} \otimes \bold{D} +\bold{B} \otimes \bold{D}.
\end{align}

\subsubsection{Property 2: (anti)commutativity} 

\begin{align}
\bold{A}\otimes \bold{B} &= -\bold{B}\otimes \bold{A}, \\
\bold{A}\odot \bold{B} &=\bold{B} \odot \bold{A}.
\end{align}

\subsubsection{Property 3: Representation of matrix multiplication} 

\begin{equation}
\bold{AB} = \bold{A}\odot \bold{B} + i \bold{A} \otimes \bold{B}.
\end{equation}
c.f. $(\bold{a}\cdot \ {\bm{\mathrm{\sigma}}})(\bold{b}\cdot  {\bm{\mathrm{\sigma}}})=\bold{a}\cdot \bold{b} + i (\bold{a}\times \bold{b})\cdot  {\bm{\mathrm{\sigma}}}$ for the spin case.

\subsubsection{Property 4: Traces}
Definition: $\text{Tr}_\text{v} [\bold{A}]\doteq\text{Tr}[A]$.

\begin{align}
\text{Tr}_\text{v}[\bold{A}\otimes \bold{B}] &= 0,\\
\text{Tr}_\text{v} [\bold{A}\odot \bold{B}] &= 2 \bold{A}\cdot \bold{B}.
\end{align}

\subsubsection{Property 5: Completeness relation.}

\begin{equation}
\text{Tr}_v [\bold{A}\odot \bold{B}] = \frac{1}{2}\text{Tr}_\text{v} [\bold{A}\odot L_i]\text{Tr}_\text{v} [L_i \odot \bold{B}].
\end{equation}

\subsubsection{Property 6: Trace of matrix multiplication.}
\begin{align}
\text{Tr}[AB]&= 2 \bold{A}\cdot \bold{B},\\
\text{Tr}[ABC]&= 2 \bold{A} \cdot (\bold{B}\odot \bold{C} + i \bold{B}\otimes \bold{C}),
\end{align}
where $\bold{A}\cdot \bold{B}=A_iB_i$.

\subsubsection{Property 7: Jacobi identities}
\begin{align}
\bold{A}\otimes (\bold{B}\otimes \bold{C}) + \bold{B} \otimes (\bold{C}\otimes \bold{A})+ \bold{C} \otimes (\bold{A}\otimes \bold{B} ) &= 0,\\
\bold{A}\otimes (\bold{B}\odot \bold{C}) + \bold{B} \otimes (\bold{C}\odot \bold{A})+ \bold{C} \otimes (\bold{A}\odot \bold{B} ) &= 0.
\end{align}

\subsubsection{Property 8: Non-associativity relations}
\begin{align}
\bold{A}\odot (\bold{B}\odot \bold{C})-(\bold{A}\odot \bold{B})\odot \bold{C} &= -\bold{B}\otimes (\bold{C}\otimes \bold{A}),\\
\bold{A}\otimes (\bold{B}\otimes \bold{C}) -(\bold{A}\otimes \bold{B})\otimes \bold{C} &= -\bold{B} \otimes (\bold{C}\otimes \bold{A}),\\
\bold{A}\otimes (\bold{B}\odot \bold{C}) -(\bold{A}\otimes \bold{B})\odot \bold{C} &= -\bold{B} \odot (\bold{C}\otimes \bold{A}).
\end{align} 
c.f. The second relation is equivalent to the first Jacobi identity.

\subsubsection{Property 9: Triple scalar products}
\begin{align}
\bold{A}\cdot (\bold{B}\odot \bold{C}) &= (\bold{A}\odot \bold{B}) \cdot \bold{C} \ \ (=\frac{1}{4}\text{Tr}[ABC+ACB]), \\
\bold{A}\cdot (\bold{B}\otimes \bold{C}) &= (\bold{A}\otimes \bold{B}) \cdot \bold{C} \ \ (=\frac{1}{4i}\text{Tr}[ABC-ACB]).
\end{align}
c.f. $\bold{A}\cdot (\bold{B}\times \bold{C}) = \bold{(A\times B)\cdot C}$ for usual cross product.

\subsubsection{Property 10: spin limit}
For  $L_i = \sigma_i \ (i=0,1,2,3)$,
\begin{align}
\bold{A\odot B} \ &\dot{=} \ (\bold{A}\cdot \bold{B}, A_0B_1+B_0A_1, A_0B_2+A_2B_0,A_0B_3+A_3B_0),\\
\bold{A\otimes B} \ &\dot{=} \ (0,A_2B_3-A_3B_2,A_3B_1-A_1B_3,A_1B_2-A_2B_1 ).
\end{align}

\subsubsection{Property 11: Equation of motion}
For physical operator $A$ and Hamiltonian $H$,
\begin{equation}
\frac{dA}{dt}=\frac{1}{i\hbar}[H,A]\ \dot{=}\ \frac{2}{\hbar}\bold{H\otimes A}.  
\end{equation}

\subsection{The Green's function formalism for orbital pumping}

In this section, we derive a pumping formula applicable to arbitrary orbital systems with angular momentum quantum number $l$. We extend the formula derived in the reference~\cite{chen2015} for spin pumping, utilizing the Green's function approach, to operate even in states with higher angular momentum $(l\geq 1/2)$. For state with total angular momentum $l$, we need $(2l+1)^2 $-dimensional vector space to completely describe pumping effects. We define unit vector in $(2l+1)^2$ space as follows,
\begin{equation}
     {\bm{\mathrm{ {\bm{\mathrm{\hat{\alpha}}}}}}}\ \dot{=} \ L_\alpha.
\end{equation}
Then $\nu$-direction total current is given by,
\begin{equation}
\bold{j}_\nu  = \frac{\hbar}{4\pi}\frac{J_{ex}\hbar^2}{m_e i}\int dr' \left[ \bold{g}^R(r,r'){\bm{\mathrm{ {\bm{\mathrm{\hat{\alpha}}}}}}} \lrd \bold{g}^A(r',r)\right]\frac{du_\alpha}{dt},
\end{equation}
where $\bold{g}^{R/A}(r,r')$ is retarded and advanced Green's function defined in $(2l+1)^2$-dimensional vector space and $(du_\alpha/dt) {\bm{\mathrm{ {\bm{\mathrm{\hat{\alpha}}}}}}}$ constitutes external perturbations also defined in $(2l+1)^2$-dimensional vector space. Then, we use orbital algebra introduced in the previous section to simplify above equation. Using generalized products defined in previous section, the orbital current is given by,
\begin{align}
\bold{j}_\nu &= \frac{\hbar}{4\pi}\frac{J_{ex}\hbar^2}{m_e}\Im\int dr' \times \nonumber\\
\quad & \left[ \Big(\bold{g}^R(r,r')\odot {\bm{\mathrm{ {\bm{\mathrm{\hat{\alpha}}}}}}}\Big)\odot \lrd\bold{g}^A(r',r) -\Big(\bold{g}^R(r,r')\otimes {\bm{\mathrm{ {\bm{\mathrm{\hat{\alpha}}}}}}}\Big)\otimes \lrd\bold{g}^A(r',r)  +i\Big(\bold{g}^R(r,r') \odot {\bm{\mathrm{\hat{\alpha}}}}\Big)\otimes \lrd\bold{g}^A(r',r) + i\Big(\bold{g}^R(r,r')\otimes  {\bm{\mathrm{\hat{\alpha}}}}\Big)\odot \lrd\bold{g}^A(r',r) \right] \frac{du_\alpha}{dt}.
\end{align} 
By projecting the expression for $\bold{g}^{R/A}$ onto unit vectors, we can obtain pumping expressions with a generalized mixing conductance,
\begin{align} \label{generalcurrent}
\bold{j}_\nu&=\frac{\hbar}{4\pi}\frac{J_{ex}\hbar^2}{m_e }\int dr' \times\nonumber\\
&\sum_{\beta\gamma}  \text{Im}[g^R_\beta(r,r') \lrd g^A_\gamma(r',r)] [( {\bm{\mathrm{ {\bm{\mathrm{\hat{\beta}}}}}}}\odot  {\bm{\mathrm{\hat{\alpha}}}})\odot  {\bm{\mathrm{\hat{\gamma}}}} -( {\bm{\mathrm{\hat{\beta}}}}\otimes  {\bm{\mathrm{\hat{\alpha}}}})\otimes  {\bm{\mathrm{ {\bm{\mathrm{\hat{\gamma}}}}}}}] + \text{Re}[g^R_\beta(r,r') \lrd g^A_\gamma(r',r)][ ( {\bm{\mathrm{\hat{\beta}}}} \odot  {\bm{\mathrm{\hat{\alpha}}}})\otimes  {\bm{\mathrm{\hat{\gamma}}}}+  ( {\bm{\mathrm{\hat{\beta}}}} \otimes  {\bm{\mathrm{\hat{\alpha}}}})\odot  {\bm{\mathrm{\hat{\gamma}}}} ]\frac{du_\alpha}{dt},
\end{align}
where $g^{R/A}_m = \bold{g}^{R/A}\cdot \vhat{m}$.

\section{Orbital pumping based on the scattering matrix approach}

In this subsection, we show that the pumped orbital current by the magnetization dynamics using the scattering matrix approach~\cite{tserkovnyak2002} is consistent with that of the Green's function approach. Following the conventional scattering matrix approach~\cite{tserkovnyak2002}, the pumped current density operator $j_\alpha$ where $\alpha$ is the flow direction is given by,
\begin{align}
j_\alpha/(-e)&=\frac{\partial n_\alpha}{\partial X}\frac{dX(t)}{dt}, \\
\frac{\partial n_\alpha}{\partial X} &= \frac{1}{4\pi i} \frac{\partial S}{\partial X}S^\dagger +{\rm h.c.},
\end{align}   
where $X(t)$ is a time-dependent system parameter, $\partial n_\alpha/\partial X$ is $3\times 3$ emissivity matrix along $\alpha$ direction, and $S$ is the $3\times 3$ scattering matrix in $p$-orbital space. We then define the orbital mixing conductance as the orbital counterpart of the spin mixing conductance,
\begin{align}
    G^{i,j}\doteq 1-r_{i}r_{j}^*,
\end{align}
where $r_{i}$ and $r_j$ are the reflection coefficients. Now we calculate the pumped orbital currents by the magnetization dynamics where the Hamiltonian is given by, 
\begin{equation}\label{scattering}
    H(t) =\bold{L} \cdot \bold{M}(t).
\end{equation}
For this case, scattering matrix is given by,
\begin{eqnarray}
S = \sum_m r_{m}\ket {m}\bra {m},
\end{eqnarray}
where $\ket{m}$ is the eigenstate of the Eq.~\eqref{scattering}. Then, the pumped OAM and OAP currents are given by,
\begin{align}
    \textbf{j}^{\text{OAM}}_\alpha &= \frac{1}{8\pi}\Re\left[\left(\frac{G^{1,0}+G^{0,-1}}{2}\right)\left(\textbf{M}\times\frac{d\textbf{M}}{dt}-i\frac{d\textbf{M}}{dt}\right)\right],\label{OAM_1}\\
    j^{\text{OAP}}_{\alpha, \beta\gamma} &=\frac{1}{16\pi}\Re\left[\left(\frac{G^{1,0}-G^{0,-1}}{2}\right)\right.\nonumber\\
    &\quad\quad\quad\quad\times\left.\left(M_\beta\left(\textbf{M}\times \frac{d\textbf{M}}{dt}\right)_\gamma-iM_\beta\frac{dM_\gamma}{dt}+(\beta\leftrightarrow \gamma) \right)\right], \label{OAP_1}
\end{align}
where $\alpha$ is the surface-normal direction. We can see there is direct mapping between pumped orbital currents calculated by Green's function approach. The same conclusion can be drawn for the orbital pumping by the lattice dynamics.

\section{Linear response calculation}

We consider currents carrying a physical quantity $Q$. An increase of $Q$ in the region less than the $(\ell+1)$-th layer is given by,
\begin{align}
\frac{dN^Q(\ell)}{dt} =\frac{d}{dt}\left\langle \sum_{i=-\infty}^\ell Q^{(i)}(t) \right\rangle,
~~~~Q^{(i)}(t)=\sum_{\alpha,\alpha'} C^{\dagger}_{i\alpha}(t) Q_{i\alpha ,i\alpha'} C_{i\alpha'}(t). \nonumber
\end{align}
From the Heisenberg equation of motion, the increasing rate of $Q$ is expressed as,
\begin{align}
\frac{ d N^Q(\ell) }{dt} = \frac{1}{i\hbar} \left\langle \left[ \sum_{i=-\infty}^\ell Q^{(i)}(t),\sum C_{i_1\alpha_1}^\dagger H^{i_1,i_2}_{\alpha_1,\alpha_2}C_{i_2\alpha_2 } \right] \right\rangle, \nonumber
\end{align}
where $H$ is the Hamiltonian matrix.
Using the commutation relations of
$\{C_{i\alpha },C^\dagger_{j\alpha' }\}=\delta_{ij}\delta_{\alpha\alpha'}$,
we derive an increase of $Q$ in terms of the generation and transfer terms:
\begin{align}
\frac{ d N^Q(\ell) }{dt}  = W^Q_L(\ell) + I^Q_L(\ell). \nonumber
\end{align}
Here, $W^Q_L(\ell)$ is the generation rate and
$I^Q_L(\ell)$ is the current of increasing $Q$ in the region less than $(\ell+1)$,
\begin{align}
I^Q_L(\ell) &= \frac{1}{i\hbar} Q_{\ell\alpha,\ell\alpha'}\left\langle C_{\ell\alpha}^\dagger H_{\alpha',\alpha_1}^{\ell,\ell+1} C_{\ell+1\alpha_1}-C_{\ell+1\alpha_1 }^\dagger H_{\alpha_1 ,\alpha}^{\ell+1,\ell} C_{\ell\alpha'}  \right\rangle. \nonumber
\end{align}
By defining the lesser Green function by $\langle C^\dagger_\alpha C_\beta \rangle = -i\hbar G^<_{\beta\alpha}$,
we arrive at
\begin{align}
I^Q_L(\ell) &= 2\Re\Tr \left[ H^{\ell+1,\ell} Q_{\ell,\ell} G^{<(\ell,\ell+1)} \right]. \nonumber
\end{align}
In a similar way, we can calculate an increase of $Q$ in the region greater than the $\ell$-th layer that is given by,
\begin{align}
I^Q_R(\ell) &= 2\Re\Tr \left[ H^{\ell,\ell+1} Q_{\ell+1,\ell+1} G^{<(\ell+1,\ell)} \right]. \nonumber
\end{align}
In an averaged sense, we define the current carrying $Q$ as,
\begin{align}
j^{Q}(\ell)&= \frac{1}{2}\Big[ I_R^Q(\ell-1) -I_L^Q(\ell)  \Big]\nonumber\\
&= \Re\Tr \left[ Q_{\ell,\ell} \left( G^{<(\ell,\ell-1)} H^{\ell-1,\ell}-G^{<(\ell,\ell+1)} H^{\ell+1,\ell} \right) \right].
\label{ISymm}
\end{align}

If a time-dependent term $U(t)$ in the Hamiltonian is slow enough relative to electron dynamics, we can treat its time-derivative ${\dot U}(t)$ as a perturbation.
Then, the lesser Green function is expressed as
\begin{align}
G^<(t,t) &= -\frac{1}{2\pi\hbar} \int dE  f_0(E)\left[ G^C(E)\!+i\hbar\left(\frac{\partial G^R}{\partial E} {\dot U}(t) G^C(E)
\!-\!G^C(E) {\dot U}(t) \frac{\partial G^A}{\partial E} \right)\!+\!\cdots \right]. \nonumber
\end{align}
Here, $f_0(E)$ is the Fermi-Dirac distribution and $G^C(E)\equiv G^R(E)-G^A(E)$.
$G^{R,A}(E)=g^{R,A}[E-U(t)]$ is the retarded and advanced Green function, respectively, 
and we express them as the adiabatic modulation of the unperturbed  Green functions, $g^{R,A}(E)=(E-H\pm i\Gamma)^{-1}$ for given level broadening $\Gamma=25~{\rm meV}$.
Because the zeroth order term is irrelevant to the pumping currents,
we omit it and obtain the lesser Green function up to the first order as

\[ \delta G^<(t,t) =
-\frac{1}{i\hbar} \int dE \left( 
 \frac{\partial f_0}{\partial E} \delta\mathcal{N}_{\rm surf}(E)+ f_0(E) \delta\mathcal{N}_{\rm sea}(E) \right),  \]
where $\delta\mathcal{N}_{\rm surf}(E)$ and $\delta\mathcal{N}_{\rm sea}(E)$ are Fermi surface and sea contributions, respectively, given by~\cite{bonbien2020} 
\begin{align}
\delta\mathcal{N}_{\rm surf}(E) &= \frac{\hbar}{4\pi} \left[ G^C{\dot U}(t) G^C \right], \nonumber\\
\delta\mathcal{N}_{\rm sea}(E) &= \frac{\hbar}{4\pi} \left[ G^C{\dot U}(t)\left( \frac{\partial G^R}{\partial E}+\frac{\partial G^A}{\partial E} \right) + {\rm h.c.} \right]. \nonumber
\end{align}
By associating the lesser Green function and the currents of Eq. (\ref{ISymm}), we obtain
\begin{align}
j^Q(\ell) &= \frac{1}{\hbar} {\rm ImTr} \int dE \left[ Q_{\ell,\ell} \left\{ \frac{\partial f_0}{\partial E} \left( \delta\mathcal{N}_{\rm surf}^{\ell,\ell+1}H^{\ell+1,\ell}-\delta\mathcal{N}_{\rm surf}^{\ell,\ell-1}H^{\ell-1,\ell} \right) + f_0(E) \left( \delta\mathcal{N}_{\rm sea}^{\ell,\ell+1}H^{\ell+1,\ell}-\delta\mathcal{N}_{\rm sea}^{\ell,\ell-1}H^{\ell-1,\ell} \right) \right\} \right].
\label{Iresponse}
\end{align}

\section{Tight-binding model}

Consider a three-dimensional NM1/NM2/NM1 system of simple cubic structure (Fig.~1 in the main text). The tight-binding Hamiltonian composed of $s$ and $p$ orbitals is given as~\cite{slater1954}
\begin{align}
H^{(0)}({\bf k}) &=
\begin{pmatrix}
\ddots & \\
& \mathcal{T}^{(0)\dagger}_{11} & h^{(0)}_{1} & \mathcal{T}^{(0)}_{12} & 0 & 0 & 0 & 0 & \\
& 0 & \mathcal{T}^{(0)\dagger}_{12} & h^{(0)}_{2} & \mathcal{T}^{(0)}_{22} & 0 & 0 & 0 & \\
& & & & \ddots & & & & & \\
& 0 & 0 & 0 & \mathcal{T}^{(0)\dagger}_{22} & h^{(0)}_{2} & \mathcal{T}^{(0)}_{21} & 0 & \\
& 0 & 0 & 0 & 0 & \mathcal{T}^{(0)\dagger}_{21} & h^{(0)}_{1} & \mathcal{T}^{(0)}_{11} & &  \\
& & & & & & & & \ddots
\end{pmatrix},
\end{align}
where
\begin{align}
h^{(0)}_i - \mathcal{E}^{(0)}_i &= 
\begin{pmatrix}
2V^{(0)}_{ss\sigma}(c_x+c_y) & 2iV^{(0)}_{sp\sigma}s_x & 2iV^{(0)}_{sp\sigma}s_y & 0 \\
-2iV^{(0)}_{sp\sigma}s_x & 2V^{(0)}_{pp\sigma}c_x+2V^{(0)}_{pp\pi}c_y & 0 & 0 \\
-2iV^{(0)}_{sp\sigma}s_y & 0 & 2V^{(0)}_{pp\pi}c_x+2V^{(0)}_{pp\sigma}c_y & 0 \\
0 & 0 & 0 & 2V^{(0)}_{pp\pi}(c_x+c_y)
\end{pmatrix}, \nonumber \\
\mathcal{E}^{(0)}_i &=
\begin{pmatrix}
\epsilon_s^{(0)} & 0 & 0 & 0 \\
0 & \epsilon_{p_x}^{(0)} & 0 & 0 \\
0 & 0 & \epsilon_{p_y}^{(0)} & 0 \\
0 & 0 & 0 & \epsilon_{p_z}^{(0)}
\end{pmatrix},
\qquad
\mathcal{T}^{(0)}_{ij} = 
\begin{pmatrix}
V^{(0)}_{ss\sigma} & 0 & 0 & V^{(0)}_{sp\sigma} \\
0 & V^{(0)}_{pp\pi} & 0 & 0 \\
0 & 0 & V^{(0)}_{pp\pi} & 0 \\
-V^{(0)}_{sp\sigma} & 0 & 0 & V^{(0)}_{pp\sigma}
\end{pmatrix}, \nonumber
\end{align}
with abbreviations $c_{x(y)}\equiv\cos k_{x(y)}a_0$ and $s_{x(y)}\equiv\sin k_{x(y)}a_0$ are used for brevity. 
The basis of $h^{(0)}_i$ and $\mathcal{T}^{(0)}_{ij}$ $(i,j=1,2)$ is $\{|s\rangle,|p_x\rangle,|p_y\rangle,|p_z\rangle\}$ where subscripts are indices for the layers, NM1 or NM2.  The tight-binding parameters are set as $\epsilon_s^{(0)}=0.50$, $\epsilon_{p_x}^{(0)}=\epsilon_{p_y}^{(0)}=\epsilon_{p_z}^{(0)}=-0.70$, $V_{ss\sigma}^{(0)}=-0.30$, $V_{pp\sigma}^{(0)}=0.50$, $V_{pp\pi}^{(0)}=-0.20$, and $V_{sp\sigma}^{(0)}=0.40$ for both NM1 and NM2 layers, all in units of eV. As a side remark, the $sp$ hybridization is introduced here to create orbital texture [right panel of Fig.~\ref{fig:s1}(b)], which is important for orbital transport (i.e., the mixing between $p_x$ and $p_y$ orbitals through $sp$ hybridization). If the next-nearest neighbor hopping is considered, orbital texture can be formed through the direct mixing of $p_x$ and $p_y$ orbitals without $sp$ hybridization, though the two cases differ only in the coefficients of the effective $p$-orbital Hamiltonian~\cite{han2023}.

To impose the boundary condition that both NM1 layers are semi-infinite, we utilize the repetitive structure of Hamiltonian~\cite{luisier2006}. Namely, the Hamiltonian is constructed as
\begin{align}
H^{(0)}({\bf k}) &=
\begin{pmatrix}
\mathcal{H}_{ll} & \mathcal{I}_{la} & 0 \\
\mathcal{I}_{al} & \mathcal{H}_{aa} & \mathcal{I}_{ar} \\
0 & \mathcal{I}_{ra} & \mathcal{H}_{rr}
\end{pmatrix}, \nonumber
\end{align}
where $\mathcal{H}$ and $\mathcal{I}$ is the on-site and interaction matrices of each region: the left contact ($l$), the right contact ($r$),
and the active region ($a$). We define the retarded Green function in a similar manner, 
\begin{align}
G^R =
\begin{pmatrix}
G_{ll} & G_{la} & G_{lr} \\
G_{al} & G_{aa} & G_{ar} \\
G_{rl} & G_{ra} & G_{rr}
\end{pmatrix}, \nonumber
\end{align}
thereby obtain the Green function of the active region as
\begin{align}
\left(\mathcal{H}_{aa}-\Sigma_l-\Sigma_r\right)G^R_{aa}=1, \nonumber
\end{align}
with the surface self energies defined as
\begin{align}
\Sigma_l = \mathcal{I}_{al}\mathcal{H}^{-1}_{ll}\mathcal{I}_{la}, \quad
\Sigma_r = \mathcal{I}_{ar}\mathcal{H}^{-1}_{rr}\mathcal{I}_{ra}. \nonumber
\end{align}

\subsection{Time-dependent lattice dynamics}

Now, suppose that a time-dependent biaxial stress is applied along two axes perpendicular to each other, $\vhat{e}_1=\vhat{z}\cos\phi+\vhat{x}\sin\phi$ and $\vhat{e}_2=-\vhat{z}\sin\phi+\vhat{x}\cos\phi$, with phase difference $\varphi$. Then, the position vector of nearest neighbors is modified as
\begin{align}
{\pmb \delta}(t) = \vhat{e}_1~\delta_1^{(0)}(1+\varepsilon_1\sin\omega t) + \vhat{e}_2~\delta_2^{(0)}[1+\varepsilon_2\sin(\omega t+\varphi)], \nonumber
\end{align}
where $\varepsilon_{1(2)}$ is the maximum strain along the axis $\vhat{e}_{1(2)}$ and $\delta_{1(2)}^{(0)}=\lim_{\varepsilon\rightarrow0}{\pmb \delta}(t)\cdot\vhat{e}_{1(2)}$. Due to the deformation of crystal structure, the directional cosines for nearest-neighbor bonds and the magnitude of hopping integrals should be altered. The bond-length dependence of hopping integrals is usually fitted to a power law~\cite{froyen1979,harrison1980}
\begin{align}
V_{\alpha\beta\tau}(d) = V_{\alpha\beta\tau}(d_0) \left(\frac{d_0}{d}\right)^{\eta_{\alpha\beta\tau}}, 
\end{align}
in which $\alpha,\beta$ denote orbitals under consideration and $\tau$ is the bonding type $\sigma,\pi,\delta,\cdots$. In general, the exponent $\eta_{\alpha\beta\tau}$ is orbital- and material-specific parameter obtained by comparing tight-binding and \textit{ab initio} data. For simplicity, however, we arbitrarily choose the exponent to be a constant for all hopping integrals, i.e., $\eta_{\alpha\beta\tau}=2$.
Then, the Hamiltonian at instantaneous time $t$ is 
\begin{align}
H({\bf k};t) &= 
\begin{pmatrix}
\ddots & \\
& \mathcal{T}^{(0)\dagger}_{11} & h^{(0)}_{1} & \mathcal{T}^{(0)}_{12} & 0 & 0 & 0 & 0 & \\
& 0 & \mathcal{T}^{(0)\dagger}_{12} & h_{2}(t) &\mathcal{T}_{22}(t) & 0 & 0 & 0 & \\
& & & & \ddots & & & & & \\
& 0 & 0 & 0 & \mathcal{T}^{\dagger}_{22}(t) & h_{2}(t) & \mathcal{T}^{(0)}_{21} & 0 & \\
& 0 & 0 & 0 & 0 & \mathcal{T}^{(0)\dagger}_{21} & h^{(0)}_{1} & \mathcal{T}^{(0)}_{11} & &  \\
& & & & & & & & \ddots
\end{pmatrix}, \nonumber
\end{align}
where
\begin{align}
\langle s;{\bf k}|h_{2}(t)|s;{\bf k}\rangle &= \epsilon_s^{(0)}+2V_{ss\sigma}^{(0)}(\tilde{c}_x+c_y), \nonumber \\
\langle p_x;{\bf k}|h_{2}(t)|p_x;{\bf k}\rangle &= \epsilon_{p_x}^{(0)} +  2(V_{pp\sigma}^{(0)}l_x^2+V_{pp\pi}^{(0)}n_x^2)\tilde{c}_x + 2V_{pp\pi}^{(0)}c_y, \nonumber \\
\langle p_y;{\bf k}|h_{2}(t)|p_y;{\bf k}\rangle &= \epsilon_{p_y}^{(0)} + 2V_{pp\pi}^{(0)}\tilde{c}_x+2V_{pp\sigma}^{(0)}c_y, \nonumber \\
\langle p_z;{\bf k}|h_{2}(t)|p_z;{\bf k}\rangle &= \epsilon_{p_z}^{(0)} + 2(V_{pp\sigma}^{(0)}n_x^2+V_{pp\pi}^{(0)}l_x^2)\tilde{c}_x+2V_{pp\pi}^{(0)}c_y, \nonumber \\
\langle s;{\bf k}|h_{2}(t)|p_x;{\bf k}\rangle &= 2iV_{sp\sigma}^{(0)}l_x\tilde{s}_x, \nonumber \\
\langle s;{\bf k}|h_{2}(t)|p_y;{\bf k}\rangle &= 2iV_{sp\sigma}^{(0)}s_y, \nonumber \\
\langle s;{\bf k}|h_{2}(t)|p_z;{\bf k}\rangle &= 2iV_{sp\sigma}^{(0)}n_x\tilde{s}_x, \nonumber \\
\langle p_x;{\bf k}|h_{2}(t)|p_z;{\bf k}\rangle &= 2(V_{pp\sigma}^{(0)}-V_{pp\pi}^{(0)})l_xn_x\tilde{c}_x, \nonumber
\end{align}
and
\begin{align}
\mathcal{T}_{22}(t) &= \frac{1}{|{\bf d}_z(t)|^2}
\begin{pmatrix}
V_{ss\sigma}^{(0)} & l_zV_{sp\sigma}^{(0)} & 0 & n_zV_{sp\sigma}^{(0)} \\
-l_zV_{sp\sigma}^{(0)} & l_z^2V_{pp\sigma}^{(0)}+n_z^2V_{pp\pi}^{(0)} & 0 & l_zn_z(V_{pp\sigma}^{(0)}-V_{pp\pi}^{(0)}) \\
0 & 0 & V_{pp\pi}^{(0)} & 0 \\
-n_zV_{sp\sigma}^{(0)} & l_zn_z(V_{pp\sigma}^{(0)}-V_{pp\pi}^{(0)}) & 0 & n_z^2V_{pp\sigma}^{(0)}+l_z^2V_{pp\pi}^{(0)}
\end{pmatrix}. \nonumber
\end{align}
Here, we neglect the time-dependent variation of on-site energies and assume $\varepsilon_1=\varepsilon_2=\varepsilon$. Note that the strain is uniformly applied in the active region, i.e., $\partial\varepsilon/\partial y=0$. The position vectors directing nearest neighbors are
\begin{align}
{\bf d}_z(t) &= a_0\vhat{z} [ 1+\varepsilon\{ \cos^2\phi\sin\omega t+\sin^2\phi\sin(\omega t+\varphi)\} ] + a_0\vhat{x}~\varepsilon\cos\phi\sin\phi[\sin\omega t-\sin(\omega t+\varphi)], \nonumber \\
{\bf d}_x(t) &= a_0\vhat{z}~\varepsilon\cos\phi\sin\phi[\sin\omega t-\sin(\omega t+\varphi)] + a_0\vhat{x} [ 1+\varepsilon\{ \sin^2\phi\sin\omega t+\cos^2\phi\sin(\omega t+\varphi) \} ], \nonumber
\end{align}
and corresponding directional cosines are defined as $l_i=\vhat{x}\cdot{\bf d}_i(t)/|{\bf d}_i(t)|$ and $n_i=\vhat{z}\cdot{\bf d}_i(t)/|{\bf d}_i(t)|$ for $i=x,z$. The revised abbreviations $\tilde{c}_{x}\equiv\cos k_{x}a_0/|{\bf d}_{x}(t)|^2$ and $\tilde{s}_{x}\equiv\sin k_{x}a_0/|{\bf d}_{x}(t)|^2$ are used.
Up to the linear order of strain $\varepsilon$,
\begin{align}
h_{2}(t)-h^{(0)}_{2} &\approx -2\varepsilon \left( \sin^2\phi\sin\omega t+\cos^2\phi\sin(\omega t+\varphi) \right)
\begin{pmatrix}
2V_{ss\sigma}^{(0)}c_x & 2iV_{sp\sigma}^{(0)}s_x & 0 & 0 \\
-2iV_{sp\sigma}^{(0)}s_x & 2V_{pp\sigma}^{(0)}c_x & 0 & 0 \\
0 & 0 & 2V_{pp\pi}^{(0)}c_x & 0 \\
0 & 0 & 0 & 2V_{pp\pi}^{(0)}c_x
\end{pmatrix} \nonumber \\
&\quad +\varepsilon\cos\phi\sin\phi(\sin\omega t-\sin(\omega t+\varphi))
\begin{pmatrix}
0 & 0 & 0 & 2iV_{sp\sigma}^{(0)}s_x \\
0 & 0 & 0 & 2(V_{pp\sigma}^{(0)}-V_{pp\pi}^{(0)})c_x \\
0 & 0 & 0 & 0 \\
-2iV_{sp\sigma}^{(0)}s_x & 2(V_{pp\sigma}^{(0)}-V_{pp\pi}^{(0)})c_x & 0 & 0
\end{pmatrix}, \nonumber
\end{align}
and
\begin{align}
\mathcal{T}_{22}(t)-\mathcal{T}^{(0)}_{22} &\approx -2\varepsilon \left( \cos^2\phi\sin\omega t+\sin^2\phi\sin(\omega t+\varphi) \right)
\begin{pmatrix}
V_{ss\sigma}^{(0)} & 0 & 0 & V_{sp\sigma}^{(0)} \\
0 & V_{pp\pi}^{(0)} & 0 & 0 \\
0 & 0 & V_{pp\pi}^{(0)} & 0 \\
-V_{sp\sigma}^{(0)} & 0 & 0 & V_{pp\sigma}^{(0)}
\end{pmatrix} \nonumber \\
&\quad +\varepsilon\cos\phi\sin\phi(\sin\omega t-\sin(\omega t+\varphi))
\begin{pmatrix}
0 & V_{sp\sigma}^{(0)} & 0 & 0 \\
-V_{sp\sigma}^{(0)} & 0 & 0 & V_{pp\sigma}^{(0)}-V_{pp\pi}^{(0)} \\
0 & 0 & 0 & 0 \\
0 & V_{pp\sigma}^{(0)}-V_{pp\pi}^{(0)} & 0 & 0
\end{pmatrix}. \nonumber
\end{align}

We superpose two biaxial strains to generate lattice vibration with finite net angular momentum. One biaxial strain with $\phi_1=0$ and $\varphi_1=\pi$ and the other strain with $\phi_2=\pi/4$ and $\varphi_2=\pi$ are overlapped where overall phase of the latter is shifted by $\pi/2$ with respect to the former. Note that the strength of each set of strains is set to $\varepsilon=0.5\%$. To obtain an insight on the perturbation we take, consider the tight-binding Hamiltonian of $p$ orbitals in bulk simple cubic lattice. Then, the time-dependent perturbation $U^{\rm bulk}(t)$ in the long-wavelength limit is given as
\begin{align}
U^{\rm bulk}(t) \approx -4\varepsilon(V_{pp\sigma}^{(0)}-V_{pp\pi}^{(0)})\left[ \{L_z,L_x\}\cos\omega t-(L_z^2-L_x^2)\sin\omega t\right],
\end{align}
which resembles $K[{\bf L}\cdot{\bf u}(t)]^2$ with halved frequency for $\hat{\bf u}(t)=\vhat{z}\cos\omega t+\vhat{x}\sin\omega t$. For this model, the strain-OAP coupling strength $K$ is determined by $-4\varepsilon(V_{pp\sigma}^{(0)}-V_{pp\pi}^{(0)})$. Thus, one can conclude that the effective perturbation $K[{\bf L}\cdot{\bf u}(t)]^2$ captures the lattice dynamics characterized by the circular motion of an atom about its equilibrium position, which in turn gives rise to adiabatic transformation of the OAP.

\section{Decay length of orbital currents}

\begin{figure}[t]
\begin{center}
\includegraphics[scale=0.5]{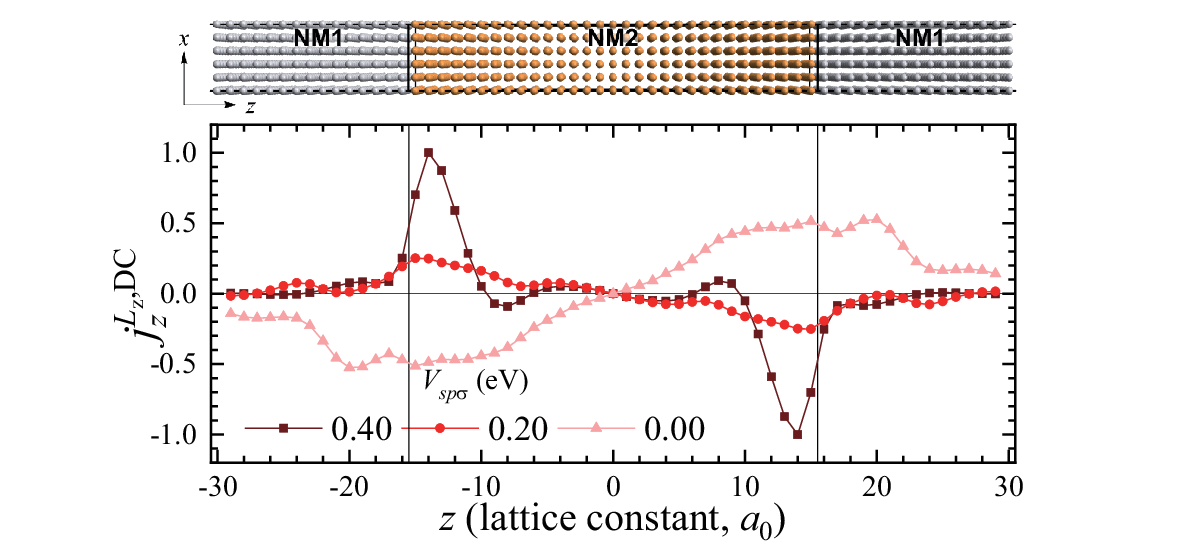}
\caption{Spatial profile of DC orbital pumping current driven by the time-dependent magnetic field ${\bf L}\cdot{\bf M}(t)$ for $\hat{\bf M}(t)=\vhat{x}\cos\omega t+\vhat{y}\sin\omega t$ for various strengths of $sp$ hybridization $V_{sp\sigma}$.
}
\label{fig:s2}
\end{center}
\end{figure}

The transmission of OAM current into adjacent layers depends on the orbital character~\cite{go2023,urazhdin2023}. For a given geometry, an electron propagating toward the leads experiences the crystal field which splits $|p_z\rangle$ from $|p_x\rangle$ and $|p_y\rangle$. Therefore, one can suppress the spatial oscillation of OAM response by choosing $\hat{\bf M}(t)=\vhat{x}\cos\omega t+\vhat{y}\sin\omega t$ that conveys the OAM $L_z$. The calculated OAM current $j^{L_z,\rm DC}_z$ (Fig.~\ref{fig:s2}) exhibits a monotonically decaying behavior rather than an oscillatory decay near the interface. Still, the OAM penetration length is not strikingly enhanced as that previously reported in ferromagnets when the orbital texture exists. Under consideration of all electrons participating in the propagation, we recognize that degenerate $|p_x\rangle$ and $|p_y\rangle$ states with nonzero in-plane wave vectors are gapped into $|p_r\rangle\equiv\cos\phi_{\bf k}|p_x\rangle+\sin\phi_{\bf k}|p_y\rangle$ and $|p_t\rangle\equiv-\sin\phi_{\bf k}|p_x\rangle+\cos\phi_{\bf k}|p_y\rangle$ by the orbital texture while the degeneracy at ${\bf k}=\pm k_F\vhat{z}$ are preserved. Note that $\phi_{\bf k}=\arg(k_x+ik_y)$ is an azimuthal angle of wave vector $\bf k$. As the operator $\hat{L}_z=i|p_y\rangle\langle p_x|-i|p_x\rangle\langle p_y|=i|p_t\rangle\langle p_r|-i|p_r\rangle\langle p_t|$, the broken degeneracy between $|p_r\rangle$ and $|p_t\rangle$ results in increased oscillation for states carrying $L_z$, and thus short-ranged orbital transport. Conversely, the overall penetration length can be extended by decreasing the strength of orbital texture as shown in Fig.~\ref{fig:s2}.

\begin{figure}[b]
\begin{center}
\includegraphics[scale=0.5]{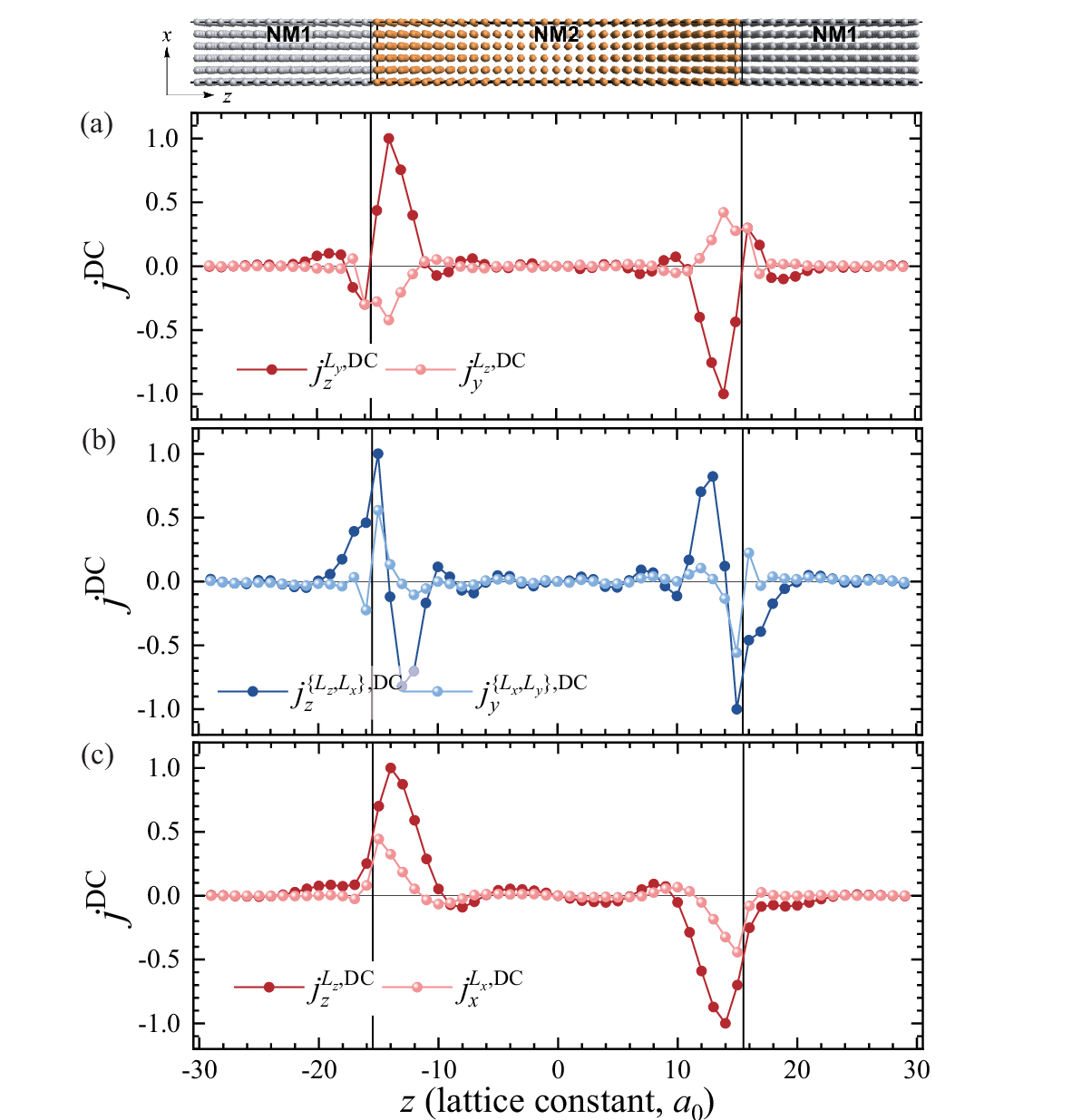}
\caption{Spatial profile of DC orbital pumping currents and resultant orbital swapping currents driven by the time-dependent magnetic field ${\bf L}\cdot{\bf M}(t)$ for (a,b) $\hat{\bf M}(t)=\vhat{z}\cos\omega t+\vhat{x}\sin\omega t$ and (c) $\hat{\bf M}(t)=\vhat{x}\cos\omega t+\vhat{y}\sin\omega t$.
}
\label{fig:s3}
\end{center}
\end{figure}

\section{Orbital swapping effect}

The pumped orbital current serves as a primary input of the orbital swapping effect~\cite{manchon2023}, which is an orbital counterpart of the spin swapping effect~\cite{lifshits2009,sadjina2012,saidaoui2016}. Similar to the spin swapping effect which is categorized into two types, the orbital swapping effect can be classified into one that illustrates the conversion of OAM current $j_{\alpha,\beta}^{\rm OAM}\rightarrow j_{\beta,\alpha}^{\rm OAM}$ for $\alpha\ne \beta$ [Fig.~\ref{fig:s3}(a)] and the other which represents the conversion of OAM current $j_{\alpha,\alpha}^{\rm OAM}\rightarrow j_{\beta,\beta}^{\rm OAM}$ for $\alpha\ne \beta$ [Fig.~\ref{fig:s3}(c)]. Furthermore, we observe the swapping of OAP current $j^{\rm OAP}_{\alpha,\alpha\beta}\rightarrow j^{\rm OAP}_{\gamma,\beta\gamma}$ [Fig.~\ref{fig:s3}(b)] and $j^{\rm OAP}_{\alpha,\beta\gamma}\rightarrow j^{\rm OAP}_{\beta,\gamma\alpha}$ (not shown). Note that the latter process converts AC OAP current to another AC OAP current in our model. All processes are suppressed when the orbital texture vanishes.

\section{Effect of inversion symmetry breaking}

In this section, we investigate the effect of inversion symmetry breaking (ISB) in orbital pumping. ISB is crucial in orbital transport as a finite orbital moment is allowed at each crystal momentum~\cite{bhowal2020,sahu2021}. Considering the significance of ISB in previous studies, an explicit analysis of the impact of ISB on orbital pumping is beyond the scope of our work and requires an independent study. Instead of a complete systematic study, we perform an analytic calculation on damping-like component of OAM current with ISB. The influence on other pumped currents can be inferred from this result. As a side remark, in our numerical calculation, ISB is implicitly contained as electrons traverse the interfaces of the heterostucture in which ISB takes place. 

To incorporate the effect of ISB on orbital pumping, we conducted an analytical calculation with the orbital Rashba Hamiltonian at the interface, following the methodology proposed by Ref.~\cite{chen2015}. We hypothesized that the inversion symmetry is broken along the $z$-axis,
 \begin{equation}
     H_\text{R} = \lambda_{\text{R}} (\bold{k}\times \bold{L}) \cdot \hat{\bold{z}},
 \end{equation}
where $\lambda_{\text{R}}$ is the orbital-Rashba coefficient. Building on this premise, we calculated the additional term appearing in the damping-like OAM current, up to the second order of Rashba strength. As a consequence, we find that the second order correction of OAM current has anisotropic term along $z$ direction, i.e., the direction of ISB, which is given by,
\begin{align}
    j_z^{(2),L_i}&= \left[\alpha + \beta (\bold{M}\cdot \hat{\bold{z}})^2\right]\left(\bold{M}\times\frac{d\bold{M}}{dt} \right)_i + \gamma \bold{M}\times\left[\hat{\bold{z}}\times\left(\hat{\bold{z}}\times\frac{d\bold{M}}{dt} \right) \right]_i, \\
    j_z^{(2),L_z}&=\left[\delta + \epsilon(\bold{M}\cdot \hat{\bold{z}})^2 \right]\left(\bold{M}\times\frac{d\bold{M}}{dt} \right)_z,
\end{align}
where $i$ runs for $x$ and $y$ and detailed coefficients are given below. This is a natural form in light of ISB direction ($z$-axis). A similar analysis for the spin counterpart was conducted in Ref.~\cite{chen2015,kim2012}. The terms proportional to $\alpha$ and $\delta$ serve as analogues to those reported in spin physics; the interface Rashba Hamiltonian induces rotation of the OAM about the axis $\bold{k}\times\hat{\bold{z}}$~\cite{chen2015}. Due to the direction of the Rashba term, the in-plane component and the out-of-plane OAM experience different torques ($\alpha\neq \delta$). However, the term represented by $\beta$, $\gamma$, and $\eta$ lacks a spin counterpart; it arises from the effect of the OAP to OAM conversion channel via the Rashba field. This can be indirectly observed from the broader precession channel presented in Eq.~\eqref{generalcurrent}. This suggests that the interface impact on orbital transport differs from that on spin and is more complex and implies that there is conversion channel between OAP and OAM. Nevertheless, a detailed analysis of this conversion physics at the interface falls beyond the scope of our paper.

Here, we present the form of $\alpha$, $\beta$, $\gamma$, $\delta$, and $\epsilon$. We define, $g^{\text{r/a}}_n\doteq(g^{\text{r/a}}_1-g^{\text{r/a}}_{-1})/2$ and $g^{\text{r/a}}_p\doteq(g^{\text{r/a}}_1+g^{\text{r/a}}_{-1}-2g_0^{\text{r/a}})/2$. We also define functions which are given by,
\begin{align}
    G(n,m,l,k)&\doteq \frac{J_{ex}\hbar^2(\lambda_\text{R}k_{\text{F}})^2}{16\pi m_e} \left[g^r_n(z,0)g^r_m(0,z')\overset{\leftrightarrow}{\partial_z} g_l^a(z',0)g_k^a(0,z)\right],\nonumber\\
    H(n,m,l,k)&\doteq \frac{J_{ex}\hbar^2(\lambda_\text{R}k_{\text{F}})^2}{16\pi m_e}\left[g^r_n(z,0)g^r_m(0,0)g^r_l(0,z')\overset{\leftrightarrow}{\partial_z} g_k^a(z',z)\right].
\end{align}
Then coefficients are given by,
\begin{align}
    \alpha=\text{Re}[&G(0,p,0,n)+G(p,0,n,0)+G(n,n,n,0)+G(n,n,0,n)+H(0,n,n,n)+H(n,n,0,n) \nonumber\\
    &-2H(n,n,n,0)+H(0,0,n,p)+2H(0,n,0,p)+H(n,0,0,p)+H(0,0,p,n)-H(0,n,p,0)\nonumber\\
    &-H(n,0,p,0)+H(0,n,p,p,)-H(0,p,n,0)-H(n,p,0,0)+H(0,p,n,p)+H(n,p,0,p)\nonumber\\
    &+H(0,p,p,n)-2H(n,p,p,0)+H(p,0,0,n)-H(p,0,n,0)-H(p,n,0,0)+H(p,n,0,p)\nonumber\\
    &-2H(p,n,p,0)+H(p,p,0,n)-2H(p,p,n,0)],\nonumber\\
    \beta =\text{Re}[&G(p,0,p,n)+G(0,p,0,n)+G(p,0,n,0)-G(p,n,0,p)-G(p,p,n,0)-G(p,p,0,n) \nonumber\\
&-2G(n,p,0,0)-2G(p,n,0,0)-G(n,p,p,0)-G(p,n,p,0)-G(n,n,n,0)-G(n,n,0,n) \nonumber\\
&-2H(0,0,0,n)+3H(0,n,0,0)+2H(0,n,0,0)+3H(n,0,0,0)-2H(0,p,0,n)+2H(n,n,n,0) \nonumber\\
&+2H(0,n,p,0)+3H(n,0,p,0)+2H(0,p,n,0)+2H(n,p,0,0)+2H(n,p,p,0)+3H(p,0,n,0) \nonumber\\
&+2H(p,n,0,0)+2H(p,n,p,0)+2H(p,p,n,0)], \nonumber\\
\gamma =\text{Re}[&G(p,p,n,0)+G(p,p,0,n)+G(n,p,p,0)+G(p,n,p,0)+2G(n,n,n,0)+2G(n,n,0,n) \nonumber\\
&-G(p,0,p,n)+G(p,n,0,p)+2G(n,p,0,0)+2G(p,n,0,0)-2H(0,0,0,n)-H(0,0,n,0) \nonumber\\
&-2H(0,n,0,0)-H(0,n,n,n)-H(n,0,0,0)-H(n,n,0,n)-H(0,0,n,p)-H(0,n,0,p) \nonumber\\
&-H(n,0,0,p)-H(0,0,p,n)-H(0,n,p,0)-H(0,n,p,p)-2H(0,p,0,n)-H(0,p,n,0) \nonumber\\
&-H(n,p,0,0)-H(0,p,n,p) -H(n,p,0,p)-H(n,p,0,0)], \nonumber\\
\delta = \text{Re}[&G(0,n,0,0)+G(n,0,0,0)-G(0,p,0,n)-G(p,0,n,0)-2H(0,0,0,n)+3H(0,0,n,0) \nonumber\\
&+2H(0,n,0,0)+3H(n,0,0,0)+2H(n,n,n,0)+2H(0,n,p,0)+3H(n,0,p,0)-2H(0,p,0,n) \nonumber\\
&+2H(0,p,n,0)+2H(n,p,0,0)+2H(n,p,p,0)+3H(p,0,n,0)+2H(p,n,0,0)+2H(p,n,p,0) \nonumber\\
&+2H(p,p,n,0)], \nonumber\\
\epsilon= \text{Re}[&G(p,0,p,n)-G(p,n,0,p)-G(p,p,n,0)-G(p,p,0,n)-G(n,n,n,0)-G(n,n,0,n) \nonumber\\
&-G(0,p,0,n)-G(p,0,n,0)-2G(n,p,0,0)-2G(p,n,0,0)-G(n,p,p,0)-G(p,n,p,0) \nonumber\\
&+H(0,n,n,n)+H(n,n,0,n)-2H(n,n,n,0)+H(0,0,n,p)+2H(0,n,0,p)+H(n,0,0,p) \nonumber\\
&+H(0,0,p,n)-H(0,n,p,0)-H(n,0,p,0)+H(0,n,p,p)-H(0,p,n,0)-H(n,p,0,0) \nonumber\\
&+H(0,p,n,p)+H(n,p,0,p)+H(0,p,p,n)-2H(n,p,p,0)+H(p,0,0,n)-H(p,0,n,0) \nonumber\\
&-H(p,n,0,0)+H(p,n,0,p)-2H(p,n,p,0)+H(p,p,0,n)-2H(p,p,n,0)].
\end{align}

\section{Electric quadrupole pumping}

In this section, we examine whether the second harmonics pumping of OAP remains valid even when it goes beyond the atom-center approximation. For this purpose, we use the fact that OAP is proportional to the electric quadrupole (EQ). This relation can be understood recalling that the OAP captures the angular distribution of electron distribution near atoms. For instance, $\{L_x,L_y\}$ captures whether the angular distribution follows that of $p_x+p_y$ orbital (negative contribution to $\{L_x,L_y\}$) or that of $p_x-p_y$ orbital (positive contribution to $\{L_x,L_y\}$), which exhibits the electron angular distribution rotated by $\pm$45 degree within $xy$ plane with respect to the $x$ axis. This illustrates that $\{L_x,L_y\}$ is proportional to the EQ $Q_{xy}\sim{~}xy$. Then, we calculate OAP pumping using the modern theory formulation of EQ, which includes the inter-site contribution, presented in the Ref.~\cite{daido2020} and compare it with the results of our paper. However, this formulation cannot be applied to non-periodic systems such as heterostructures. Therefore, we adopted an alternative approach based on the modern theory (but still with limited quantitative prediction power) and checked whether the qualitative features in our main text are valid. To be specific, we followed the approach suggested in Ref.~\cite{go2023orbital}. In this study, the authors calculated the time-dependent OAM density for a homogeneous bulk sample (instead of heterostructures), and assumed that its time derivative is proportional to the pumped orbital current. Similarly, we calculated the time-dependent EQ and assumed that its time derivative is proportional to the pumped EQ current.

To address the simplest case, we solved two-dimensional $p$-orbital system in which a magnetic field rotates in-plane with frequency $\omega$. We treated the effect of the magnetic field as a perturbation, and approximated up to second order, and calculated the EQ responses. As pointed out in Ref.~\cite{han2023}, a two-dimensional $p$-orbital system typically exhibits $|p_r\rangle$, $|p_t\rangle$, and $|p_z\rangle$  orbitals as eigenstates, where $|p_r\rangle\equiv\cos\phi_{\bf k}|p_x\rangle+\sin\phi_{\bf k}|p_y\rangle$ and $|p_t\rangle\equiv-\sin\phi_{\bf k}|p_x\rangle+\cos\phi_{\bf k}|p_y\rangle$, and $\phi_{\bf k}=\arg(k_x+ik_y)$. For the sake of simplicity, we focused exclusively on the $|p_t\rangle$ band, the one with the lowest energy level. Consequently, we identified two distinct EQ responses with 2$\omega$ frequency,
\begin{align}
    \frac{1}{2}(Q_{xx}-Q_{yy})&=\frac{\lambda^2}{8\pi}\int dk \frac{1}{k}\left[\frac{f(E_t)}{2(E_t-E_z)^2}+\frac{1}{(E_t-E_z)^3}\int^{\infty}_{E_t}d\epsilon f(\epsilon) \right]\cos 2\omega t, \\
    Q_{xy} &=\frac{\lambda^2}{8\pi}\int dk \frac{1}{k}\left[\frac{f(E_t)}{2(E_t-E_z)^2}+\frac{1}{(E_r-E_z)^3}\int^{\infty}_{E_t}d\epsilon f(\epsilon) \right]\sin 2\omega t,
\end{align}
where $\lambda$ represents orbital Zeeman coupling strength and $f(E)$ is the Fermi-Dirac distribution function. This corresponds to bulk density of EQ and corresponding pumped EQ currents are proportional to $d(Q_{ij})/dt$ which are given by,
\begin{align}
    j_z^{(Q_{xx}-Q_{yy})/2}  &\propto -\frac{\omega \lambda^2}{4\pi}\int dk \frac{1}{k}\left[\frac{f(E_t)}{2(E_t-E_z)^2}+\frac{1}{(E_t-E_z)^3}\int^{\infty}_{E_t}d\epsilon f(\epsilon) \right]\sin 2\omega t, \\
    j_z^{Q_{xy}} &\propto \frac{\omega \lambda^2}{4\pi}\int dk \frac{1}{k}\left[\frac{f(E_t)}{2(E_t-E_z)^2}+\frac{1}{(E_t-E_z)^3}\int^{\infty}_{E_t}d\epsilon f(\epsilon) \right]\cos 2\omega t.
\end{align}
To draw a comparison with our results obtained using OAP operators, we evaluated the pumped OAP current employing the formula detailed in the main text and we get, 
\begin{equation}
    j_z^{\text{OAP}} = \frac{\omega}{4\pi}\text{Im}[G^{-}_L]\left(\{L_x,L_y\}\cos2\omega t-(L_x^2-L_y^2)\sin 2\omega t \right),
\end{equation}
where we only write down $2\omega$ terms. Considering $Q_{xy}$ is proportional to $\{L_x,L_y\}$ while $(Q_{xx}-Q_{yy})/2$ is proportional to $(L_x^2-L_y^2)$, two results show same angular dependence. Thus, we expect that the second harmonics pumping property remains valid even beyond the atom-center approximation.

%